\documentclass[acmsmall]{acmart}

\usepackage[utf8]{inputenc}

\usepackage{amsmath}
\makeatletter
\@ifundefined{Bbbk}{}{}
\makeatother
\usepackage{amssymb}

\usepackage{bbm}
\usepackage{textcomp}
\DeclareUnicodeCharacter{2212}{-}

\usepackage{graphicx}
\usepackage{adjustbox}
\usepackage{wrapfig}

\usepackage{booktabs}
\usepackage{tabularx}
\usepackage{array}
\usepackage{multirow}
\usepackage{longtable}
\usepackage{makecell}
\usepackage{colortbl}

\usepackage{subcaption}
\usepackage{algorithm}
\usepackage{algpseudocode}

\usepackage{enumitem}

\usepackage{pgf}
\usepackage{tikz}
\usepackage{forest}
\usepackage{xspace}

\usepackage[most]{tcolorbox}
\usepackage{csquotes}
\usepackage[noorphans,vskip=1em,leftmargin=1em]{quoting}

\usepackage{xurl}

\usepackage{siunitx}
\usepackage{listings}

\newcommand{\circnum}[1]{%
  \tikz[baseline=(n.base)]{
    \node (n) [
      circle,
      fill=black,
      text=white,
      font=\bfseries\scriptsize,
      inner sep=1.2pt
    ] {#1};
  }%
}

\definecolor{PhaseColor}{RGB}{120,0,60}
\colorlet{LineNoColor}{black!55}

\algrenewcommand\algorithmicrequire{\textbf{Require:}}
\algrenewcommand\algorithmicensure{\textbf{Ensure:}}

\makeatletter
\algrenewcommand\alglinenumber[1]{%
  \scriptsize\color{LineNoColor}{#1}%
}
\@ifundefined{alglinenumberseprule}{}{%
  \algrenewcommand\alglinenumberseprule{}%
}
\makeatother

\newcommand{\Phase}[1]{%
  \State \textcolor{PhaseColor}{\#~#1}%
}

\algrenewcommand\algorithmiccomment[1]{%
  \hfill\textcolor{PhaseColor}{\#~#1}%
}

\definecolor{lightpurple}{RGB}{235,230,250}
\definecolor{grey}{gray}{0.95}

\lstdefinestyle{promptstyle}{
  basicstyle=\footnotesize\ttfamily,
  breaklines=true,
  columns=fullflexible,
  backgroundcolor=\color{gray!10},
  frame=single,
  rulecolor=\color{gray!60},
  keepspaces=true,
}

\setcopyright{cc}
\setcctype{by}
\acmDOI{10.1145/3832173}
\acmYear{2026}
\acmJournal{PACMSE}
\acmVolume{3}
\acmNumber{ISSTA}
\acmArticle{ISSTA082}
\acmMonth{10}
\acmSubmissionID{issta26main-p757-p}
\received{2026-01-29}
\received[accepted]{2026-06-25}

\begin{document}

\title[MoT: Modularization-of-Thought Prompting...]
{MoT: Modularization-of-Thought Prompting for Effective Code Generation}

\author{Ruwei Pan}
\orcid{0009-0005-1340-7242}
\affiliation{%
  \institution{Chongqing University}
  \city{Chongqing}
  \country{China}
}
\email{panruwei@stu.cqu.edu.cn}

\author{Hongyu Zhang}
\orcid{0000-0002-3063-9425}
\affiliation{%
  \institution{Chongqing University}
  \city{Chongqing}
  \country{China}
}
\email{hyzhang@cqu.edu.cn}

\renewcommand{\shortauthors}{Ruwei Pan and Hongyu Zhang}

\begin{abstract}
Large Language Models are transforming software development by automatically
generating code. Current prompting techniques such as Chain-of-Thought (CoT)
suggest tasks step by step and the reasoning process follows a linear structure,
which hampers the understanding of complex programming problems, particularly
those requiring hierarchical solutions. Inspired by the principle of
modularization in software development, in this work, we propose a novel
prompting technique called MoT (Modularization of Thought), to enhance the code
generation performance of LLMs. First, MoT exploits modularization principles
to decompose complex programming problems into smaller, independent reasoning
steps, enabling a more structured and interpretable problem-solving process.
This hierarchical structure improves the LLMs' ability to comprehend complex
programming problems. Then, it structures the reasoning process using an MLR
Graph (Multi-Level Reasoning Graph), which hierarchically organizes reasoning
steps. This approach enhances modular understanding and ensures better
alignment between reasoning steps and the generated code, significantly
improving code generation performance. Our experiments on two advanced LLMs
(GPT-4o-mini and DeepSeek-R1), comparing MoT to six baseline prompting
techniques across eight benchmarks, demonstrate that MoT significantly
outperforms existing baselines (e.g., CoT and SCoT), achieving Pass@1 scores
ranging from 58.1\% to 95.1\%.
\end{abstract}



\maketitle


\section{Introduction}

Large Language Models (LLMs) are transforming the field of software development \citep{austin2021program, athiwaratkun2022multi, brown2020language, jiang2025agentic, deng2025nocode}. In recent years, an increasing number of LLMs have been developed to assist programmers in writing code, such as GPT-4 \citep{openai2024gpt4technicalreport} and DeepSeek \citep{deepseekai2024deepseekllmscalingopensource}. 

\begin{figure}[htbp]
    \centering
    \includegraphics[width=0.8\textwidth]{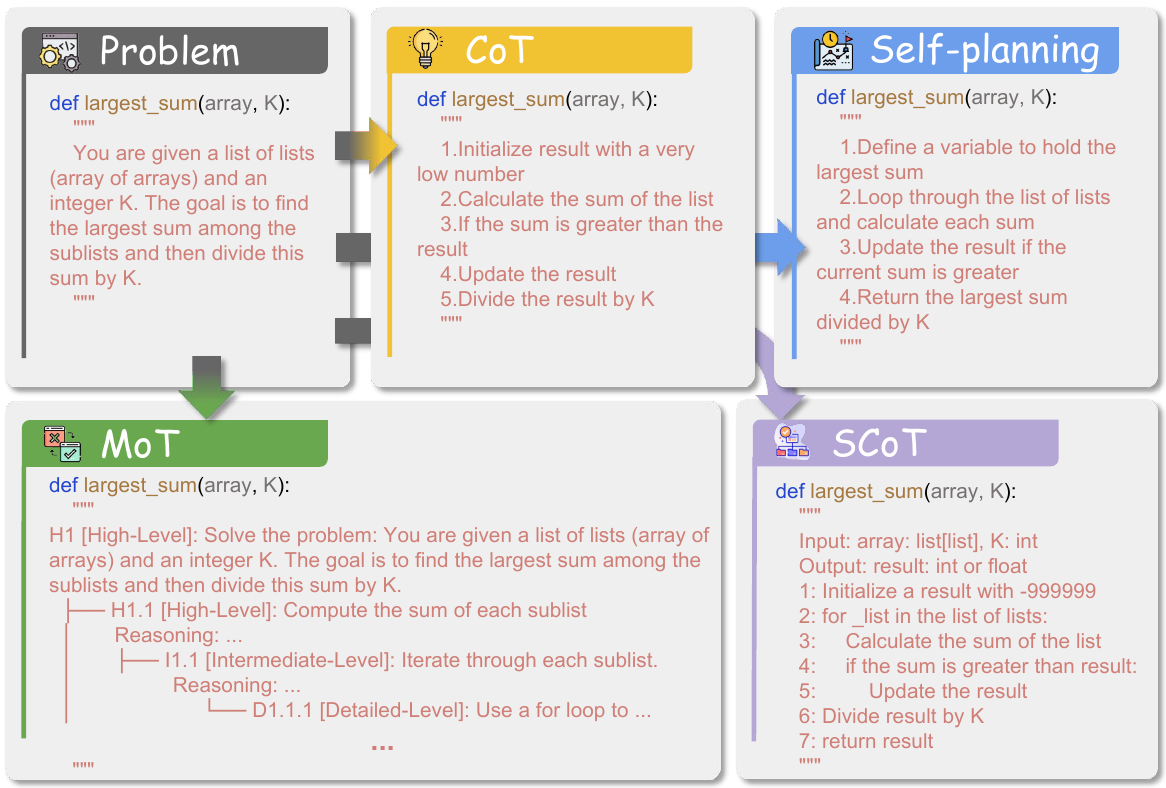}
    \caption{Different Prompting Techniques}

    \label{fig:prompt_technique}
\end{figure}

Various prompting techniques have also been introduced to enhance LLM-based code generation, as shown in Figure \ref{fig:prompt_technique}.
For example, Chain-of-Thought (CoT) prompting \citep{wei2022chain} is one of the most widely used prompting techniques, which employs intermediate steps to facilitate step-by-step reasoning in LLMs. CoT has been shown to significantly improve LLMs' problem-solving capabilities without requiring modifications to the model itself. To further enhance the quality of code generation, Li et al. \citep{li2025structured} proposed Structured Chain-of-Thought (SCoT), which builds upon CoT prompting by leveraging program structures (i.e., sequence, branch, and loop structures) to generate intermediate reasoning steps. Jiang et al. \citep{jiang2024self} introduced Self-Planning prompting, which enables LLMs to design a structured step-by-step plan prior to code generation. The LLM creates a comprehensive plan for code generation based on the problem description, which it then executes incrementally. Recently, CodeCoT prompting \citep{huang2023codecot} has integrated CoT with self-examination. The LLM improves the accuracy of the generated code by producing initial code and test cases, verifying code execution, and iteratively resolving errors as they are identified. 
Although existing prompting techniques have been proposed to guide LLMs, they still encounter several limitations due to their monolithic reasoning structure \citep{pan2025codecor, austin2021program}. 
This rigid structure limits LLMs' capacity to effectively break down complex programming problems into modular and independently solvable subproblems \citep{wei2022emergent}.

Analogous to the concept of monolithic programming in traditional software development, prompting techniques such as CoT suggest tasks step by step through a monolithic structure \citep{wang2022self, zhou2022least}. Before 1970, software development primarily adhered to a linear, monolithic programming paradigm \citep{parnas1972criteria}. 
Although this monolithic approach was relatively effective at the time, its limitations gradually became evident as software size and complexity increased, leading to reduced development efficiency.
To address the limitations of the monolithic approach, David Parnas introduced the concept of modularization in software development and highlighted the significance of hierarchical design \citep{parnas1972criteria, parnas1978buzzword, parnas2002secret}. 
He noted that by breaking complex problems into smaller, more manageable parts, developers can design and implement systems at a higher level of abstraction. Similarly, while explicit CoT methods have demonstrated effectiveness in specific tasks, their inherently linear structure limits their capacity to handle complex tasks. 
Researchers believe that the principles of modularization and hierarchical structure should also be incorporated into the reasoning processes of LLMs \citep{fowler2018refactoring}.

In this paper, we propose the MoT (\textbf{M}odularization \textbf{o}f \textbf{T}hought) prompting technique for code generation. 
MoT goes beyond traditional monolithic reasoning by introducing a structured approach that better captures the modular nature of programming problems.
At its core, MoT constructs an MLR Graph (\textbf{M}ulti-\textbf{L}evel \textbf{R}easoning Graph),
where each node represents a code task design, and each edge defines the logical dependencies between tasks.
By providing an explicit structure of relationships and functionalities, the MLR Graph enhances LLMs’ understanding of modular code components.
Inspired by the principle of modularization in software development, MoT decomposes complex programming problems into smaller, independent yet interdependent modules.
These modules can be refined, adapted, and recombined, allowing for flexible composition of a coherent final solution.
Through this modular and hierarchical approach, MoT enables LLMs to provide better solutions to programming problems.

We conduct extensive experiments to evaluate MoT with two state-of-the-art LLMs (i.e., GPT-4o-mini \citep{openai2024gpt4technicalreport} and DeepSeek-R1 \citep{deepseekai2024deepseekllmscalingopensource}) on eight benchmarks, including six function-level benchmarks (HumanEval~\citep{chen2021evaluating}, HumanEval-ET~\citep{dong2023codescore}, HumanEval+~\citep{evalplus}, MBPP~\citep{austin2021program}, MBPP-ET~\citep{dong2023codescore}, and MBPP+~\citep{evalplus}), one recent contest-style code generation benchmark (LiveCodeBench~\citep{jain2024livecodebench}), and one repository-level benchmark (RAL-Bench~\citep{pan2026ral}). 
Our results show that MoT significantly outperforms all the compared prompting techniques with both LLMs on all benchmarks. 
For example, MoT achieves an average improvement of 0.03\% to 32.85\% in Pass@1 across all subjects compared to other techniques. The results show the effectiveness of MoT for improving code generation performance. 
Moreover, we construct two variants of MoT for the ablation study. The results confirm the contribution of the modularization principles and the MLR Graph prompting strategy to the overall MoT performance. 
We also conduct an experiment to analyze cost implications, demonstrating that MoT provides an effective approach to code generation.

We summarize our contributions in this work as follows:
\begin{itemize} 
\item We propose a novel prompting technique, called MoT, to improve the code generation performance of LLMs by incorporating the modularization principles of software development into the reasoning process. 

\item  We design a novel Multi-Level Reasoning Graph to enhance modular understanding and ensure better alignment between reasoning steps and the generated code.

\item We conduct extensive experiments on two LLMs (i.e., GPT-4o-mini and DeepSeek-R1) with eight benchmarks, comparing them with six baselines to demonstrate the effectiveness of MoT in improving code generation performance. 

\end{itemize}

\section{Related Work}

In recent years, LLMs have demonstrated significant potential in the field of code generation \citep{jain2022jigsaw, li2021editsum, li2023skcoder, lu2022reacc, luo2023wizardcoder, parvez2021retrieval}. Standard language models perform code completion and generation after autoregressive pre-training. Academia and industry have introduced various code LLMs, such as AlphaCode \citep{li2022alphacode}, CodeGen \citep{nijkamp2022codegen}, CodeGeeX \citep{zheng2023codegeex}, InCoder \citep{fried2022incoder}, StarCoder \citep{li2023starcoder}, CodeLlama \citep{rozière2024codellamaopenfoundation}, and CodeT5+ \citep{Zheng2023codet5}. Furthermore, general LLMs, such as ChatGPT \citep{openai2024gpt4technicalreport} and DeepSeek \citep{deepseekai2024deepseekllmscalingopensource}, are also widely used for code generation. 
Building on these advances in model capability, recent studies have investigated how prompting techniques and structured reasoning can further improve code generation performance.

Prompting techniques have driven the 
advancement of LLM-based code generation 
\citep{chang2024efficient}. 
Basic prompting techniques include zero-shot and few-shot prompting.
Zero-shot prompting generates code based solely on the problem description without any examples, while few-shot prompting aids the LLM in understanding the input-output structure by incorporating examples \citep{chen2021evaluating}. 
Beyond these basic prompting techniques, representative prompting techniques include CoT \citep{wei2022chain}, Self-Planning \citep{jiang2024self}, and SCoT \citep{li2025structured}, which guide LLMs through explicit intermediate reasoning before code generation. CodeCoT \citep{huang2023codecot} further combines reasoning with self-examination and iterative repair.

In addition to these prompting techniques, prior work has also explored structured reasoning methods beyond monolithic reasoning. Tree-of-Thoughts~\citep{yao2023tree} and Graph-of-Thoughts~\citep{besta2024graph} extend reasoning from a single linear chain to tree- or graph-structured exploration over multiple intermediate thoughts, typically involving exploration and selection among alternative reasoning paths. More closely related to code generation, Parsel~\citep{zelikman2023parsel} decomposes problems into hierarchical function descriptions for compositional program synthesis, while CodeChain~\citep{le2023codechain} and CodeTree~\citep{li2025codetree} further improve code generation through iterative self-revision or tree-structured refinement under execution or agent feedback. These methods show the benefits of structured reasoning, but they differ from MoT in their primary mechanisms: ToT and GoT emphasize multi-path exploration, Parsel emphasizes compositional function specifications, and CodeChain and CodeTree emphasize iterative refinement or feedback-driven repair. 
In contrast, 
MoT focuses on the planning stage before final code generation by constructing a fixed-depth Multi-Level Reasoning graph as a lightweight planning scaffold to guide a single plan-to-code generation process.
This design makes hierarchical decomposition directly usable for downstream code generation without introducing explicit search, backtracking, or iterative repair.

\section{Approach}

\subsection{Overview}

\begin{figure*}[htbp]
    \centering
    \includegraphics[width=\textwidth]{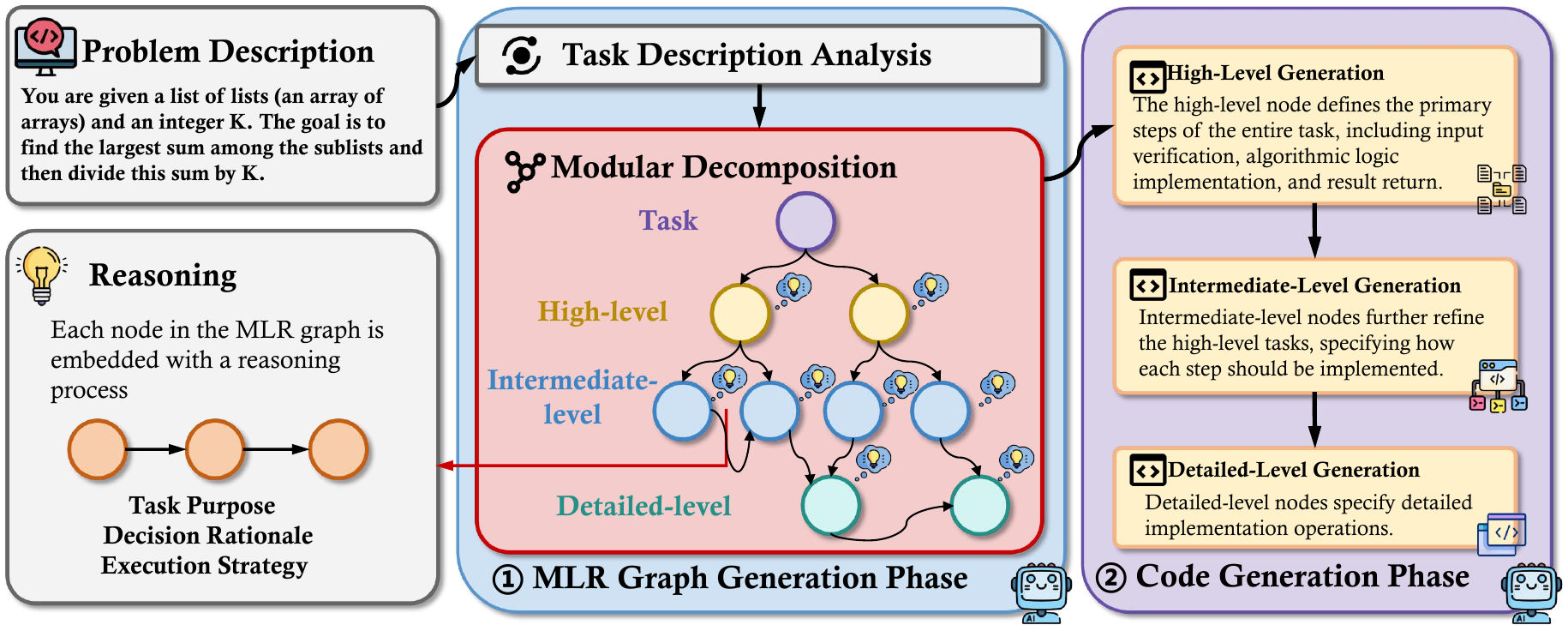}
 
    \caption{The Overview of MoT.}
    \label{fig:overview}
    \vspace{-10px}

\end{figure*}

We introduce MoT (\textbf{M}odularization-\textbf{o}f-\textbf{T}hought Prompting), a novel prompting technique based on MLR graphs (\textbf{M}ulti-\textbf{L}evel \textbf{R}easoning), to enhance LLMs' code generation. MoT applies modularization and hierarchical structures to guide logical reasoning and code generation using MLR graphs.
Our core idea is to incorporate the modularization principles into the reasoning process. This novel Multi-Level Reasoning graph can improve modular understanding and ensure better alignment between 
reasoning steps and the generated code.

Figure \ref{fig:overview} provides an overview of our method, which comprises two primary phases: (1) the MLR graph generation phase (see Section 3.2 for details), where the model analyzes the problem description and constructs an MLR graph of the programming problem; (2) the code generation phase (see Section 3.3 for details), where the model progressively generates code guided by the hierarchical MLR graph, thereby enhancing code accuracy through the modularization and hierarchical structure. The prompts that we used in our approach are shown in our repository.

\subsection{MLR Graph Generation Phase}

\begin{figure*}[htbp]
    \centering
    \includegraphics[width=\textwidth]{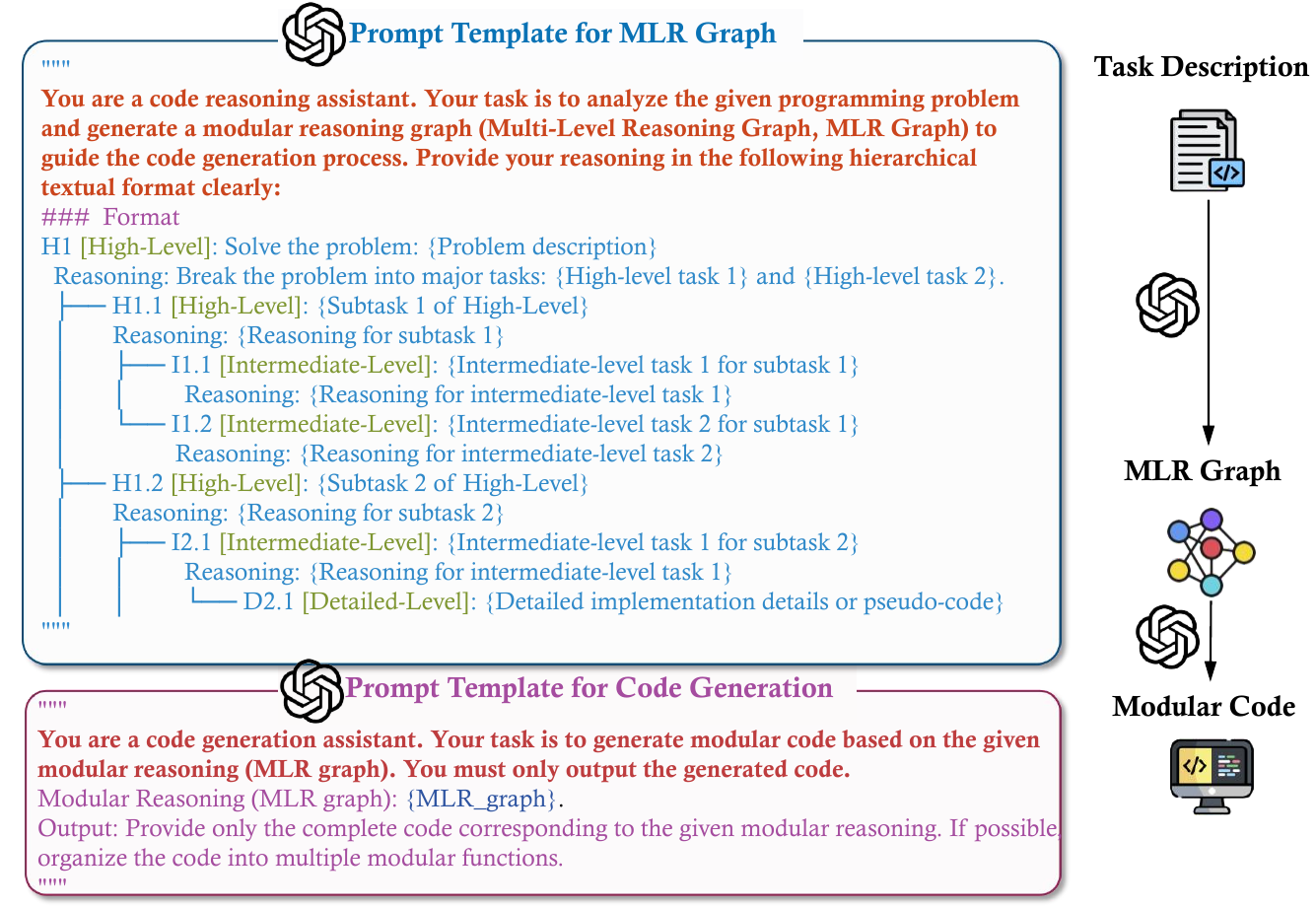}
    \caption{Prompt Templates for the MLR Graph Generation and Modular Code Generation.}

    \label{fig:prompt_code}
    \vspace{-10px}

\end{figure*}

A Multi-Level Reasoning (MLR) graph is a structured representation that systematically structures reasoning steps into prompts, facilitating LLMs' comprehension of complex programming problems. 
In an MLR graph, each node indicates the rationale for its step by embedding the reasoning process. During the MLR graph generation phase, the LLM first generates an MLR graph, categorizing high-level, intermediate-level, and detailed-level designs, according to the description of programming problems. This structured prompting provides guidance during code generation, helps improve modular understanding, and ensures better alignment between
reasoning steps and the generated code.
Figure \ref{fig:prompt_code} illustrates the prompt template (in the blue area) utilized in the MLR graph generation phase. 

\subsection{Multi-Level Reasoning (MLR) Graph}
In the MLR graph generation phase, MoT transforms task description analysis into an explicit hierarchical plan (the MLR graph) that guides subsequent code generation. 
The MLR graph uses a fixed three-level hierarchy to provide sufficient decomposition for complex tasks while preventing excessive fragmentation. Nodes represent sub-tasks at different abstraction levels, and directed edges capture decomposition-induced dependencies (a parent sub-task is refined into child sub-tasks).
MLR graphs enable the model to provide modular designs for programming problems by decomposing tasks step by step and embedding a reasoning process.

\textbf{\textit{Task description analysis: }} Firstly, the LLM extracts key task information from the programming problem, including the goal description, input and output specifications, and related constraints. The goal description typically outlines the core requirements of the programming problem. The LLM identifies relevant features based on the input specifications, such as data type, list length limits, or condition thresholds. The analysis results form the foundation for modular decomposition. 

\textbf{\textit{Modular decomposition: }}The MLR graph structures task design into three hierarchical levels:

\begin{itemize}
    \item \textbf{High-Level:} The high-level node designs the primary steps of the entire task, including input verification, algorithmic logic implementation, and result return. 
    Input validation ensures that the input data meets requirements, while the algorithmic logic implements the core steps of the task. 
    The result return node is responsible for output generation and ensuring that the final result meets expectations.
    \item \textbf{Intermediate-Level:} Intermediate-level nodes further refine the high-level tasks, specifying how each step should be implemented. For example, input validation can be broken down into subtasks such as checking for empty lists and validating data types. Algorithm implementation can be decomposed into defining loop structures and making conditional judgments.
    \item \textbf{Detailed-Level:} Detailed-level nodes specify detailed implementation operations, such as calculating numerical differences or performing conditional checks. Detailed-level nodes describe local operations needed for correctness, and they are not intended to force a separate function for every step.
\end{itemize}

Each node in an MLR graph is embedded with a reasoning process, which explains the task purpose, decision rationale, and execution strategy:

\begin{itemize}
    \item \textbf{Task Purpose: }Clarifies the need for the step, such as ensuring input validation to prevent errors in subsequent logic.
    \item \textbf{Decision Rationale: }Explains the reasoning behind specific choices, for example, using a nested loop to traverse all element combinations.
    \item \textbf{Execution Strategy: }Describes the operational details of the step, such as determining whether conditions are met or handling boundary cases.
\end{itemize}

By embedding the reasoning process, the LLMs can gain a better understanding of the design of each node in the MLR graph. 
In this way, the MLR graph not only delineates the logical structure of task decomposition but also systematically supplies the LLM with the necessary reasoning information for task execution, enhancing its code generation capabilities and logical consistency when processing complex tasks. For simpler tasks or nodes, the LLM can compress this reasoning process and inline the corresponding implementation to avoid over-modularized wrappers.

\subsubsection{Shortcut Policy}

MoT adopts a simple shortcut policy during MLR graph generation to control decomposition granularity and the verbosity of the reasoning process. This policy is training-free and is implemented as explicit prompting rules.

\begin{itemize}
    \item \textbf{Stop expansion.} When generating the MLR graph level by level, the LLM may stop further expansion for a node if the sub-task is already specific enough to be implemented directly and does not benefit from additional planning. In particular, we stop expansion under the following conditions. \circnum{1} The sub-task is directly implementable, for example, it can be expressed as a simple function. \circnum{2} The sub-task does not require non-trivial control-flow design, such as multi-branch logic, loop structure decisions, or delicate boundary handling. Otherwise, the model continues expanding the sub-task to the next level (up to the fixed three-level hierarchy). This typically happens when the sub-task involves multiple constraints, requires non-trivial branching or looping, or contains boundary cases that benefit from explicit decomposition.
    
    \item \textbf{Reasoning compression.} Each node follows the same reasoning process schema: task purpose, decision rationale, and execution strategy. To reduce the risk of over-modularization, we apply the following rule. \circnum{1} We use a compressed reasoning process when the sub-task is simple and self-evident. In this case, we keep a brief statement of the task purpose and execution strategy, and omit or shorten the decision rationale. \circnum{2} We keep the full reasoning process when the sub-task involves non-trivial design choices or complex boundary handling. Typical cases include selecting control-flow structure, choosing data structures, coordinating multiple constraints, or addressing corner cases.
\end{itemize}

Therefore, MoT acts as a planning scaffold rather than a strict function-per-node design. Decomposition is used to improve planning quality, while simple nodes can be inlined and only reusable or non-trivial nodes are 
implemented as helper functions.

\subsubsection{An Example.}

\begin{figure*}[t]
    \centering
    \includegraphics[width=\textwidth]{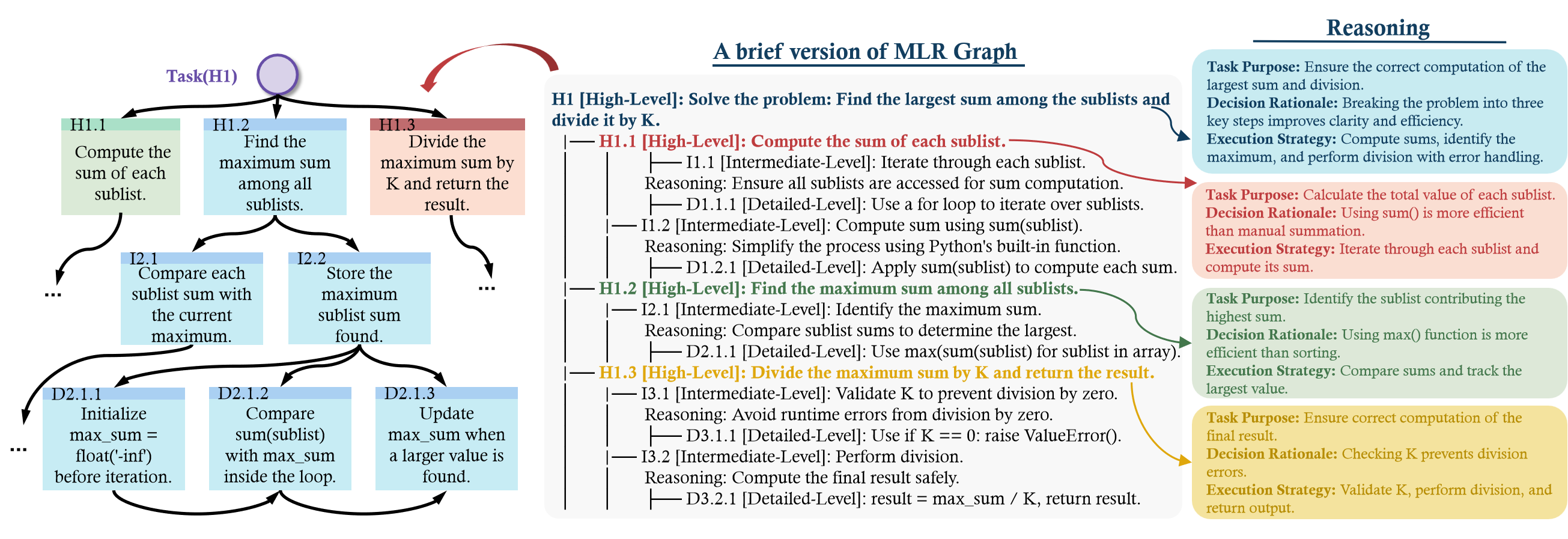}
   
    \caption{Example of MLR Graph for the problem in Figure \ref{fig:overview}.}

    \label{fig:MLR_example}
    \vspace{-8pt}
\end{figure*}

In Figure \ref{fig:MLR_example}, we present a specific example of an MLR graph generated from the programming problem description in Figure \ref{fig:overview} (partial). 
The complete MLR graph can be found in the repository.
This figure demonstrates the modular decomposition and step-by-step reasoning process enabled by the MLR graph for complex programming tasks. 
The problem is to find the largest sum among the sublists and divide it by $K$.
The MLR graph in Figure \ref{fig:MLR_example}
has three high-level (e.g., "Compute the sum of each sublist"), five intermediate-level (e.g., "Iterate through each sublist"), and five detailed-level nodes (e.g., "Use a for loop to iterate over sublist"), enabling modular reasoning.
For each node, the reasoning consists of three elements: Task Purpose, Decision Rationale, and  Execution Strategy. 
For example, the task purpose of the high-level node "Compute the sum of each sublist" is to calculate the total value of each sublist for subsequent comparison. The decision rationale of the node emphasizes the use of the Python built-in function \textit{sum()} for efficiency and simplicity. The execution strategy explicitly states that the calculation is performed by iterating over the sublist and invoking the \textit{sum()} function.

\subsection{Code Generation Phase}

In the code generation phase, the LLM leverages the generated MLR graphs to progressively produce code according to the task decomposition structure defined by the nodes.  
The code generation phase builds upon the MLR graph established in the previous phase, enabling the model to effectively complete various components of complex tasks. 
In this phase, the previously generated MLR graph guides the LLM to generate structured and modular code. The model progressively produces the implementation by following the hierarchical nodes defined in the MLR graph (from high-level to detailed-level). 
Each node guides either a reusable helper function or a local code segment, depending on its complexity and reuse potential. Simple nodes can be inlined rather than materialized as separate functions.

\begin{figure*}[htbp]
    \centering
    \includegraphics[width=\textwidth]{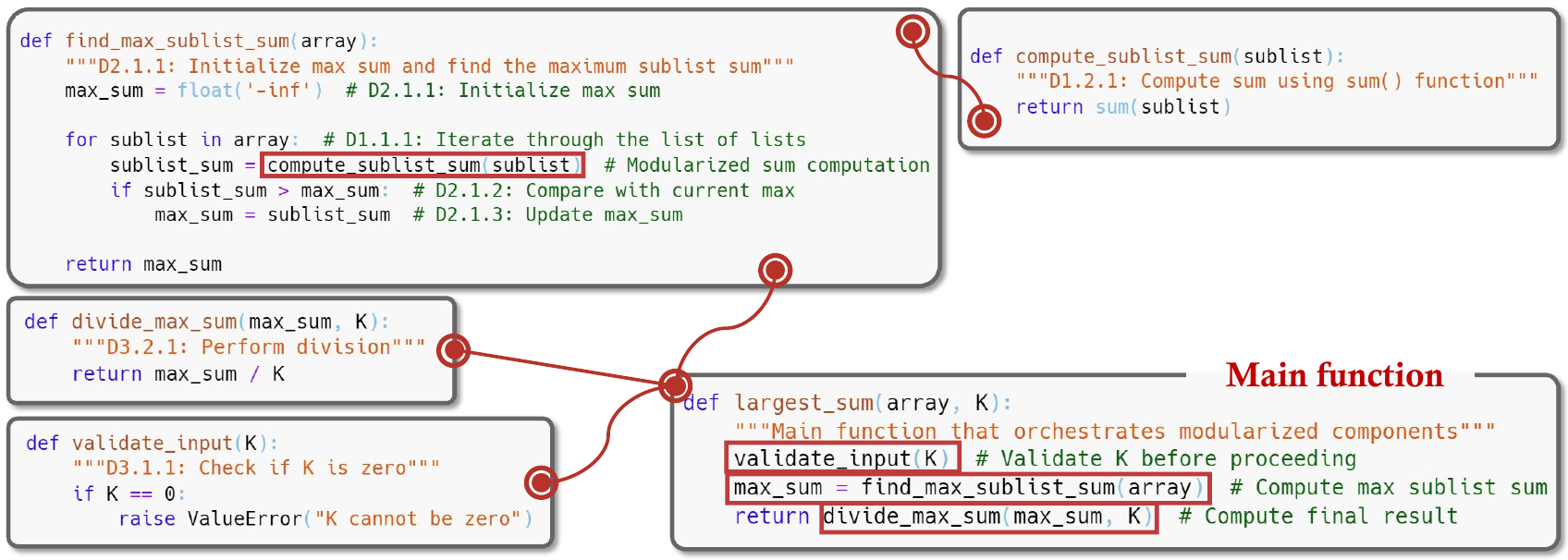}
    \caption{Example of the generated code for the problem in Figure \ref{fig:overview}.}
    \label{fig:code_example}
    \vspace{-2mm}

\end{figure*}

\subsubsection{An Example.}

To better illustrate how the MLR graph guides modular code generation, we present a generated final code based on the MLR graph in Figure \ref{fig:code_example}. The code reflects the hierarchical and modular reasoning defined in the MLR graph.

The high-level task is executed through \textit{largest\_sum()}, which serves as the primary computation function. It is responsible for calling each module sequentially to complete input verification, calculate the maximum sublist sum, and handle the final division operation. The intermediate task further refines the logic, where \textit{find\_max\_sublist\_sum()} is responsible for processing the input array and computing the maximum sublist sum, while \textit{validate\_input()} ensures the validity of K. The detailed task specifically implements the core computational steps, where \textit{compute\_sublist\_sum()} efficiently calculates the sum of a single sublist using Python's built-in \textit{sum()} function, and \textit{divide\_max\_sum()} performs the final division operation, dividing the maximum sublist sum by K to obtain the final result. The entire execution process adheres to the hierarchical design of the MLR graph, ensuring modular reasoning and implementation.
The structured guidance provided by the MLR graph can improve the comprehension of the programming problem and lead to better code.

\subsection{Overall Algorithm}

The MoT algorithm (Algorithm \ref{alg:mot}) constructs a hierarchical MLR graph to guide code generation, ensuring logical consistency and correctness. Although we use the term “MLR graph”, in our prompt it is a hierarchical framework with only parent–child decomposition edges (i.e., a 3-level DAG without same-level links). This structure justifies the nested traversal in Algorithm \ref{alg:mot}. MoT consists of two phases. 

\textit{\textbf{In Phase I (MLR Graph Generation)}}, the model parses the task description $T$ and extracts key task elements, including goal $G_T$, I/O specification $I_T$, and constraints $C_T$. It then constructs the MLR graph level by level: High-level nodes $N_T$ capture the main sub-tasks, Intermediate-level nodes $N_M$ refine each high-level sub-task into implementable steps, and Detailed-level nodes $N_B$ specify local operations and boundary handling. To reduce ambiguity, each node is associated with a reasoning process (task purpose, decision rationale, execution strategy). For simple nodes, we allow a compressed reasoning process to avoid over-modularization.

\textit{\textbf{In Phase II (Code Generation)}}, MoT first generates a code framework from the high-level nodes, including function names, signatures, and interfaces, which enforces alignment between plan nodes and code units. It then fills in the implementation by generating code snippets for detailed nodes and attaching them to the corresponding code unit. Finally, \textsc{FinalizeCode} integrates all code units into the complete output $C_f$.

\textit{Implementation note.} Algorithm \ref{alg:mot} expands the MLR graph in a depth-first manner (finish all children before moving to the next sibling). We choose depth-first expansion to preserve local context and reduce reasoning drift along a sub-task chain. The exact prompts for our method are included in our released artifact.

\begin{algorithm}[t]
\caption{Modularization-of-Thought Prompting for Code Generation}
\label{alg:mot}
\footnotesize
\begin{algorithmic}[1]
\Require Task description $T$
\Ensure Final generated code $C_f$

\Phase{\textbf{Phase-I: MLR Graph Generation}}
\State $(G_T, I_T, C_T) \gets \textsc{ExtractTaskElements}(T)$
\State $MLRG \gets \textsc{InitializeMLR}()$ \Comment{3-level hierarchy with parent--child edges}
\State $\{N_T\} \gets \textsc{GenerateHighLevel}(G_T, I_T, C_T)$
\ForAll{$N_T \in \{N_T\}$}
  \State $\textsc{EmbedReasoning}(N_T,\texttt{compact})$
  \State $\{N_M\} \gets \textsc{GenerateIntermediateLevel}(N_T)$
  \ForAll{$N_M \in \{N_M\}$}
    \State $\textsc{EmbedReasoning}(N_M,\texttt{compact})$
    \State $\{N_B\} \gets \textsc{GenerateDetailedLevel}(N_M)$
    \ForAll{$N_B \in \{N_B\}$}
      \State $\textsc{EmbedReasoning}(N_B,\texttt{full})$
    \EndFor
  \EndFor
\EndFor
\State $MLRG \gets \textsc{AssembleHierarchy}(\{N_T\},\{N_M\},\{N_B\})$
\Comment{Depth-first expansion for context consistency}

\Phase{\textbf{Phase-II: Code Generation}}
\State $C \gets \textsc{InitializeCode}()$
\State $C \gets \textsc{GenerateScaffold}(MLRG, C)$
\Comment{names, signatures, interfaces, call relations}
\ForAll{$N_T \in MLRG$}
  \ForAll{$N_M \in N_T$}
    \ForAll{$N_B \in N_M$}
      \State $C \gets \textsc{GenerateCode}(N_B, C)$
      \Comment{attach to the mapped code unit; inline if trivial}
    \EndFor
  \EndFor
\EndFor
\State $C_f \gets \textsc{FinalizeCode}(C)$
\Return $C_f$
\end{algorithmic}
\end{algorithm}

\section{Evaluation}

We evaluate MoT by answering the following research questions (RQs):

\begin{itemize}
    \item \textbf{RQ1. How does MoT prompting perform in terms of accuracy compared to baselines?}

    \item \textbf{RQ2. How effective are the major components of MoT?}

    \item \textbf{RQ3. What are the cost implications of MoT?}

    \item \textbf{RQ4. What is the impact of MLR graph depth on code generation performance?}

    \item \textbf{RQ5. Is MoT robust to ambiguities and noisy variations in prompts?}

\end{itemize}

\vspace{-6pt}

\subsection{Datasets}

To evaluate MoT, we have conducted extensive experiments on eight representative code generation datasets: HumanEval, HumanEval-ET, HumanEval+, MBPP, MBPP-ET, MBPP+, LiveCodeBench, and RAL-Bench.

\begin{itemize}
    \item \textbf{HumanEval} \citep{chen2021evaluating} is a benchmark dataset designed to assess the code generation capabilities of large language models. It consists of 164 manually crafted Python programming problems, each accompanied by corresponding test cases to verify the correctness of the generated code.
    \item \textbf{MBPP} \citep{austin2021program} is a dataset comprising diverse Python programming problems. It contains 974 code-generation tasks that cover a wide range of programming scenarios.
    \item \textbf{HumanEval-ET and MBPP-ET} \citep{dong2023codescore} are extended versions of the original HumanEval and MBPP datasets.
    \item \textbf{HumanEval+ and MBPP+} \citep{evalplus} are enhanced versions of HumanEval and MBPP, with the EvalPlus framework augmenting each problem by adding a significantly larger set of test cases—approximately 80 times more than the original dataset.
    \item \textbf{LiveCodeBench} \citep{jain2024livecodebench} is a continuously updated code generation benchmark collected from recent programming contests.
    \item \textbf{RAL-Bench} \citep{pan2026ral} is a repository-level generation benchmark. Given a natural language requirement, the model is required to generate a complete runnable repository.

\end{itemize}

\subsection{Baselines}

To evaluate MoT, we consider six typical prompting techniques for comparisons:

\begin{itemize}
    \item \textbf{Zero-shot prompting} \citep{chen2021evaluating} is a method of generating code without utilizing any code examples. The model generates code based solely on problem descriptions.
    \item \textbf{Few-shot prompting} \citep{chen2021evaluating} allows LLMs to choose a few examples for understanding the relationship between problem and code. 
    In our experiments, we adopt the 2-shot setting. 
    \item \textbf{CoT prompting} \citep{wei2022chain} addresses complex code problems through step-by-step reasoning. When the model approaches a code problem, it first produces a sequence of intermediate steps, enhancing the logical coherence and correctness of the generated code.
    \item \textbf{Self-Planning prompting} \citep{jiang2024self} enables the model to formulate a step-by-step plan prior to code generation. The model creates a comprehensive plan for code generation based on the problem description, which it then executes incrementally.
    \item \textbf{SCoT prompting} \citep{li2025structured} builds upon CoT prompting by leveraging program structures (i.e., sequence, branch, and loop structures) to generate intermediate reasoning steps. 
    \item \textbf{CodeCoT prompting} \citep{huang2023codecot} integrates CoT with self-examination mechanisms. The model improves the accuracy of generated code by producing logically coherent initial code and corresponding test cases, verifying code execution, and iteratively resolving errors as they are identified.
\end{itemize}

For CoT, SCoT, Self-Planning, and CodeCoT, to achieve fair comparisons, in our experiments, we adopt the prompts used in the respective papers.

\begin{table*}[htbp]
    \centering
    \scriptsize
    \caption{Performance comparison across multiple datasets and models}
    \resizebox{\textwidth}{!}{%
    \begin{tabular}{l|ccccccccccc}
        \toprule
        \textbf{Methods} & \multicolumn{2}{c}{\textbf{HumanEval}} & \multicolumn{2}{c}{\textbf{HumanEval+}} & \multicolumn{2}{c}{\textbf{HumanEval-ET}} & \multicolumn{2}{c}{\textbf{MBPP}} & \textbf{MBPP+} & \multicolumn{2}{c}{\textbf{MBPP-ET}}\\
        &  Pass@1 & APR & Pass@1 & APR & Pass@1 & APR & Pass@1 & APR & Pass@1 & Pass@1 & APR\\
        \midrule
        \textbf{GPT-4o-mini} \\
        \midrule
        \textbf{Zero-shot} & 88.4  & 86.2  & 81.1  & 30.8  & 87.1  & 75.0  & 59.9  & 59.7  & 47.9  & 53.6  & 52.6  \\
        \textbf{Few-shot }& 82.3  & 63.6  & 76.2  & 30.7  & 81.7  & 64.7  & 49.1  & 49.9  & 40.6  & 48.4  & 42.7  \\
        \textbf{CoT} & 87.8  & 79.2  & 82.9  & 29.4  & 87.8  & 86.5  & 61.2  & 62.4  & 48.6  & 54.1  & 53.5  \\
        \textbf{Self-Planning} & 87.2  & 90.0  & 79.9  & 30.8  & 87.1  & 74.6  & 52.1  & 54.1  & 42.4  & 48.2  & 46.9  \\
       \textbf{SCoT} & 86.6  & 93.1  & 78.7  & 30.8  & 86.0  & 74.8  & 63.9  & 63.9  & 51.4  & 55.6  & 57.2  \\
       \textbf{CodeCoT} & 83.5  & 84.8  & 73.8  & 30.8  & 82.4  & 81.0  & 55.6  & 56.6  & 40.4  & 53.3  & 48.8  \\
\rowcolor{cyan!8}
        \textbf{MoT }& \textbf{92.1} & \textbf{96.9 }& \textbf{83.5} & \textbf{30.9} & \textbf{91.5} & \textbf{88.0} & \textbf{73.9} &\textbf{79.7 }& \textbf{58.1} &\textbf{58.9} & \textbf{64.4} \\
        \midrule
        \textbf{DeepSeek-R1} \\
        \midrule
        \textbf{Zero-shot} & 93.3  & 93.1  & 87.8  & 31.0  & 92.7  & 80.4  & 69.4  & 72.4  & 57.6  & 64.0  & 58.5  \\
        \textbf{Few-shot} & 84.7  & 66.7  & 79.9  & 31.0  & 84.1  & 75.7  & 69.4  & 69.9  & 57.6  & 64.6  & 57.3  \\
        \textbf{CoT} & 92.6  & 92.6  & 88.2  & 31.0  & 73.5  & 73.5  & 59.9  & 65.4  & 44.4  & 51.9  & 55.4  \\
        \textbf{Self-Planning} & 85.4  & 88.3  & 79.3  & 30.8  & 85.3  & 72.3  & 68.4  & 69.2  & 55.4  & 65.5  & 56.4  \\
        \textbf{SCoT} & 84.8  & 81.4  & 79.3  & 30.6  & 84.1  & 72.3  & 57.9  & 60.9  & 46.9  & 61.3  & 55.2  \\
        \textbf{CodeCoT} & 66.5  & 64.9  & 60.4  & 31.3  & 65.9  & 58.3  & 69.2  & 88.7  & 56.6  & 64.5  & 74.5  \\
\rowcolor{cyan!8}
      \textbf{MoT} & \textbf{95.1 }& \textbf{95.3} & \textbf{88.4} & \textbf{31.3} &\textbf{94.5} & \textbf{82.9} & \textbf{74.9} & \textbf{90.2} &\textbf{60.4} &\textbf{68.0 }&\textbf{78.4} \\
        \bottomrule
    \end{tabular}%
    }
    
    \vspace{-10px}
    \label{tab:performance_comparison}

\end{table*}

\subsection{Evaluation Metrics}

Following existing work~\citep{jiang2024self, dong2023codescore, li2022alphacode, nijkamp2022codegen, zheng2023codegeex, li2023starcoder, fried2022incoder, Zheng2023codet5, rozière2024codellamaopenfoundation}, we use executable test cases to verify the correctness of the generated code for each programming problem. Then, we employ \textbf{Pass@1} \citep{chen2021evaluating} and \textbf{AvgPassRatio} \citep{hendrycks2021measuring} metrics to assess the performance of LLMs in code generation.

\textit{Pass@1} measures the functional correctness of the generated code.
This metric is used to evaluate the performance of generated code in test cases. Given a programming problem, an LLM generates one code instance. The problem is considered solved only if the instance passes all test cases. Pass@1 is the percentage of solved problems out of the total number of problems. The formula is as follows:

\begin{equation}
\text{Pass@1} := \mathbb{E}_{\text{Problems}} \left[ 1 - \frac{n-c}{n} \right]
\end{equation}

\textit{AvgPassRatio} measures the correctness of the generated code based on its performance across evaluation test cases. Pass@1 focuses on whether the generated code is completely correct in the test case, so we introduce AvgPassRatio to complement it. AvgPassRatio (APR) is calculated by determining the ratio of passed evaluation test cases to the total number of evaluation test cases for each problem and then averaging this ratio across all problems. Larger AvgPassRatio values indicate better code generation performance. Note that in our experiments, APR was not calculated for the MBPP+ dataset as it lacks complete test cases.

Furthermore, \textit{benchmark-specific metrics} are used for LiveCodeBench and RAL-Bench following their original evaluation protocols~\citep{jain2024livecodebench,pan2026ral}. For LiveCodeBench, we report \textbf{Pass@1} on the full \texttt{release\_v5} set. For RAL-Bench, we report \textbf{Fun.}, which is computed as the average pass rate of functional system tests across tasks.

\subsection{Model Selection}

To ensure practical and diverse model evaluation, we selected \textbf{GPT-4o-mini} and \textbf{DeepSeek-R1} (671B) because of their exceptional \textit{Pass@1} performance and cost-efficiency. As shown in Table \ref{tab:model-comparison}, these two models achieve top-tier accuracy on \textbf{HumanEval} and \textbf{HumanEval+} benchmarks while maintaining minimal API usage cost—especially GPT-4o-mini. Together, they represent two distinct paradigms: general-purpose LLMs and reasoning-specialized LLMs. Their favorable performance-to-cost ratio makes them ideal for scalable experimentation across a wide range of code generation tasks.

\begin{table}[htbp]
\centering
\small

\caption{Pass@1 score (\%) and API cost comparison of models.}

\setlength{\tabcolsep}{2pt}
\begin{tabular}{lccccc}

\toprule
\textbf{Model} & 
\makecell{\textbf{HumanEval} \\(Pass@1)} & 
\makecell{\textbf{HumanEval+} \\(Pass@1)} & 
\makecell{\textbf{Input} \textbf{Cost} \\ (\$/1M)} & 
\makecell{\textbf{Output}  \textbf{Cost} \\ (\$/1M)} \\
\midrule
\rowcolor{lightpurple}
DeepSeek-R1        & \textbf{93.3}  & \textbf{93.1}  & --     & --     \\
GPT-4 Turbo        & 90.2  & 86.6  & 10.00  & 30.00  \\
DeepSeek-V3        & 91.5  & 86.6  & --     & --     \\
Claude 3.5 Sonnet  & 88.4  & 81.7  & 3.00   & 15.00  \\
\rowcolor{lightpurple}
GPT-4o-mini        & 88.4  & 81.1  & \textbf{0.15} & \textbf{0.60} \\
\bottomrule
\end{tabular}

\label{tab:model-comparison}
\end{table}

\begin{table*}[t]
\centering
\caption{Performance comparison on LiveCodeBench and RAL-Bench under different backbone LLMs.}
\label{tab:additional_benchmarks}
\setlength{\tabcolsep}{5pt}
\renewcommand{\arraystretch}{1.08}
\scriptsize
\resizebox{0.65\textwidth}{!}{
\begin{tabular}{l|l|cc}
\toprule
\textbf{Model} 
& \textbf{Method} 
& \textbf{LiveCodeBench} 
& \textbf{RAL-Bench} \\
\cmidrule(lr){3-4}
& 
& \textbf{Pass@1} 
& \textbf{Fun.} \\
\midrule

\multirow{7}{*}{GPT-4o-mini}
& Zero-shot      & 26.7 & 12.2 \\
& Few-shot       & 27.7 & 10.6 \\
& CoT            & 33.6 & 13.5 \\
& Self-Planning  & 21.9 & 15.4 \\
& SCoT           & 27.9 & 16.1 \\
& CodeCoT        & 33.8 & 16.4 \\
& \cellcolor{cyan!8}MoT
  & \cellcolor{cyan!8}38.3
  & \cellcolor{cyan!8}25.8 \\
\midrule

\multirow{7}{*}{DeepSeek-R1}
& Zero-shot      & 64.8 & 16.2 \\
& Few-shot       & 69.0 & 37.3 \\
& CoT            & 74.2 & 36.0 \\
& Self-Planning  & 66.4 & 40.8 \\
& SCoT           & 64.7 & 44.0 \\
& CodeCoT        &  67.3    & 45.1 \\
& \cellcolor{cyan!8}MoT
  & \cellcolor{cyan!8}75.3
  & \cellcolor{cyan!8} 46.2\\
\bottomrule
\end{tabular}
}
\end{table*}

\subsection{How Does MoT Prompting Perform in Terms of Accuracy Compared to Baselines? (RQ1)}

\textbf{\textit{1) Setup: }}To address RQ1, we apply MoT and six baseline techniques to GPT-4o-mini \citep{openai2024gpt4technicalreport} and DeepSeek-R1 \citep{deepseekai2024deepseekllmscalingopensource}. GPT-4o-mini is a compact, lightweight version of the GPT model, designed to balance response quality and processing speed. DeepSeek-R1 (671B) is an open-source LLM capable of performing high-precision NLP and code generation tasks. 
We use the default temperature parameters for GPT-4o-mini and DeepSeek-R1. 
Subsequently, we measure the effectiveness of each studied prompting technique on multiple benchmark settings. For function-level evaluation, we use six widely used benchmarks, including HumanEval, HumanEval-ET, HumanEval+, MBPP, MBPP-ET, and MBPP+, and report Pass@1 and AvgPassRatio metrics. To further reduce concerns about benchmark saturation, we also evaluate all studied prompting techniques on the full set of LiveCodeBench \texttt{release\_v5}, which contains 1,055 recent programming problems, and report Pass@1 under the official evaluation protocol. In addition, we evaluate all studied prompting techniques on RAL-Bench to examine their effectiveness on repository-level generation tasks. To mitigate randomness, we run each experiment on the six function-level benchmarks 10 times and report the average results.

\textbf{\noindent\textit{2) Results: }}In this section, we compare the performance of the MoT prompting technique across multiple datasets and models against several baseline methods.
The results are provided in Table \ref{tab:performance_comparison}.
The results provided \textbf{in the Pass@1 metric} show that MoT performs well across all datasets, achieving high scores on HumanEval (92.1), HumanEval+ (83.5), HumanEval-ET (91.5), and MBPP (73.9). 
In comparison, the performance of other prompting techniques is comparatively weaker. 
For instance, Self-Planning and SCoT do not achieve the performance level of MoT on multiple datasets, with particularly poor results on the MBPP+ and MBPP-ET datasets. 
These results demonstrate that MoT has distinct advantages in generating correct code. 
The results \textbf{in the APR metric} show that MoT achieves high APR scores across all datasets, particularly on HumanEval (96.9), HumanEval+ (30.9), and MBPP (79.7).

In addition to the six function-level benchmarks, we further evaluate MoT on LiveCodeBench and RAL-Bench. 
As shown in Table~\ref{tab:additional_benchmarks}, MoT achieves the best results with both GPT-4o-mini and DeepSeek-R1. On GPT-4o-mini, MoT reaches 38.3 Pass@1 on LiveCodeBench and 25.8 Fun. on RAL-Bench, outperforming the strongest baseline by 4.5 and 9.4 points, respectively. On DeepSeek-R1, MoT also obtains the best scores, with 75.3 Pass@1 on LiveCodeBench and 46.2 Fun. on RAL-Bench. These results suggest that MoT remains effective in both recent contest-style and repository-level code generation settings.

\begin{tcolorbox}[
  enhanced,
  colback=grey,                
  colframe=teal!60!black,       
  boxrule=0.6pt,                
  arc=10pt,                     
  left=4mm,right=4mm,           
  top=2mm,bottom=2mm,
  drop shadow={black!40!white}, 
]
\textbf{\textit{Answer to RQ1:} }
\textit{
MoT consistently improves performance across the eight benchmarks. 
These results show that MoT is effective across standard function-level, recent contest-style, and repository-level code generation tasks. 
}
\end{tcolorbox}

\subsection{How Effective Are the Major Components of MoT? (RQ2)}

\begin{table*}[htbp]
    \centering
    \caption{Performance comparison with different MoT variants}
    \resizebox{\textwidth}{!}{%
    \begin{tabular}{l|ccccccccccc}
        \toprule
         \textbf{Methods} & \multicolumn{2}{c}{\textbf{HumanEval}} & \multicolumn{2}{c}{\textbf{HumanEval+}} & \multicolumn{2}{c}{\textbf{HumanEval-ET}} & \multicolumn{2}{c}{\textbf{MBPP}} & \multicolumn{1}{c}{\textbf{MBPP+}} & \multicolumn{2}{c}{\textbf{MBPP-ET}}\\
        &  Pass@1 & APR & Pass@1 & APR & Pass@1 & APR & Pass@1 & APR & Pass@1 & Pass@1 & APR\\
        \midrule        
        \textbf{GPT-4o-mini}\\
        \midrule
        \textbf{w/o MLR Graph} &85.4  &  86.5   &77.4  &30.8     &  84.3   &  79.0    &59.9       & 61.9  &  44.6  &  54.4   & 48.6    \\
        \textbf{w/o Modularization} & 82.9   &   79.0   &   78.0   &  30.9   &  79.6   & 79.0     &  62.4     & 63.7  & 48.1  &  59.1   &  51.6   \\
         \rowcolor{cyan!8}
        \textbf{MoT} &92.1& 96.9 & 83.5 & 30.9 & 91.5& 88.0 & 73.9&  79.7 & 58.1& 58.9&  64.4\\

        \midrule
        \textbf{DeepSeek-R1} \\
        \midrule
        \textbf{w/o MLR Graph} & 64.6   &  87.5    & 59.1    &  30.1   & 61.5   &   57.9   & 36.3      & 43.5  & 31.3   &  36.0   &   32.3   \\
        \textbf{w/o Modularization}  & 58.5   &  82.4    & 56.1    &  31.0   & 56.8   &  57.7   &   41.6     &  43.1 &  35.3    & 40.5   &  33.1   \\
         \rowcolor{cyan!8}
        \textbf{MoT}& 95.1& 95.3 & 88.4 & 31.3 & 94.5 & 82.9& 74.9& 90.2 & 60.4& 68.0&   78.4  \\

        \bottomrule
    \end{tabular}%
    }

    \label{tab:performance_component}
\end{table*}

\textbf{\textit{1) Setup: }}To answer RQ2, we analyze how major components of MoT (as illustrated in Figure \ref{fig:overview}) affect its effectiveness. We compare the effectiveness of MoT with the following two variants:

\begin{itemize}
    \item \textbf{w/o MLR Graph: }
    This variant removes the MLR Graph Generation (the graph structure within the red component in Figure \ref{fig:overview}), which delineates different hierarchical task levels (high-level, intermediate-level, and detailed-level tasks). It retains Modular Decomposition and Code Generation (the yellow component) to generate reasoning chains. 

    \item \textbf{w/o Modularization:} 
    This variant removes the Modular Decomposition component (the red highlighted area within component \textcircled{1} in Figure \ref{fig:overview}) and the modular Code Generation components (the yellow highlighted area within component \textcircled{2} in Figure \ref{fig:overview}). Unlike zero-shot or CoT prompting techniques, this variant still utilizes the MLR Graph's hierarchical reasoning structure (high-level, intermediate-level, and detailed-level). Instead of decomposing the task into modular subtasks for separate code generation, the LLM generates the complete final code in a single, monolithic step based on the hierarchical understanding.

\end{itemize}

The exact prompts for these variants are included in our released artifact.

\textbf{\textit{2) Results: }}We evaluate MoT and its two variants (w/o MLR Graph and w/o Modularization) on GPT-4o-mini and DeepSeek-R1. As shown in Table \ref{tab:performance_component}, MoT consistently outperforms both variants across all datasets, confirming the importance of the MLR Graph and Modularization components. 
Notably, w/o Modularization still performs better than w/o MLR Graph on several datasets. 
This suggests that the MLR Graph provides the core planning scaffold, whereas modularized generation mainly affects how this scaffold is realized in code.
In w/o MLR Graph, the model loses the explicit task decomposition and dependency structure between subproblems, while w/o Modularization still preserves the MLR graph but no longer generates code module by module according to the graph. Removing the MLR Graph leads to substantial drops in Pass@1 scores: from 92.1 to 85.4 on HumanEval (GPT-4o-mini), and from 95.1 to 64.6 on HumanEval (DeepSeek-R1), along with declines in APR, indicating the MLR Graph's key role in hierarchical task understanding.
Without Modularization, the LLM adopts a monolithic reasoning approach, which also harms performance. On GPT-4o-mini, HumanEval drops from 92.1 to 82.9 and MBPP from 73.9 to 62.4. On DeepSeek-R1, MBPP drops sharply from 74.9 to 41.6, demonstrating the value of modular code generation.
The complete MoT approach integrates both the MLR Graph and Modularization to achieve optimal performance across all tasks.

\begin{tcolorbox}[
  enhanced,
  colback=grey,                
  colframe=teal!60!black,       
  boxrule=0.6pt,                
  arc=10pt,                     
  left=4mm,right=4mm,           
  top=2mm,bottom=2mm,
  drop shadow={black!40!white}, 
]
\textbf{\textit{Answer to RQ2:} }
\textit{Our MoT prompting technique exhibits better performance than its variants, confirming the effectiveness and necessity of its major components.
}

\end{tcolorbox}

\subsection{What Are the Cost Implications of MoT? (RQ3)}

\begin{table}[htbp]
    \centering
    \scriptsize
    \caption{Average per-task cost and token consumption on GPT-4o-mini.}
    \resizebox{0.9\textwidth}{!}{%
    \begin{tabular}{l|cccccc}
        \toprule
        \multirow{2}{*}{\textbf{Methods}} & \multicolumn{3}{c}{\textbf{HumanEval}} & \multicolumn{3}{c}{\textbf{MBPP}} \\
        & \textbf{Cost (\$)} & \textbf{In-token} & \textbf{Out-token} & \textbf{Cost (\$)} & \textbf{In-token} & \textbf{Out-token} \\
        \midrule
        \textbf{Zero-shot} & 0.0028 & 119.1 & 357.6 & 0.0025 & 45.1 & 340.5 \\
        \textbf{Few-shot} & 0.0016 & 374.6 & 58.0 & 0.0022 & 578.8 & 157.0 \\
        \textbf{CoT} & 0.0047 & 504.3 & 293.9 & 0.0044 & 510.4 & 390.5 \\
        \textbf{Self-Planning} & 0.0062 & 521.6 & 364.1 & 0.0059 & 578.9 & 488.1 \\
        \textbf{SCoT} & 0.0058 & 508.8 & 384.2 & 0.0052 & 559.7 & 479.6 \\
        \textbf{CodeCoT} & 0.0071 & 535.5 & 428.7 & 0.0050 & 534.0 & 449.1 \\
         \rowcolor{cyan!8}
        \textbf{MoT} & 0.0044 & 501.3 & 282.4 & 0.0042 & 493.6 & 382.4 \\
        \bottomrule
    \end{tabular}%
    }

    \vspace{-5px}

    \label{tab:llm_cost}
\end{table}

\begin{figure}[h]
    \centering
    \begin{subfigure}[t]{0.4\textwidth}
        \centering
        \includegraphics[width=\textwidth]{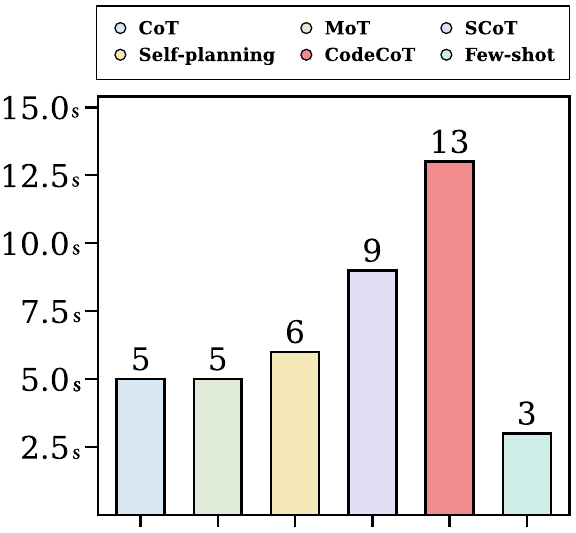}
        \caption{Average generation time on HumanEval}
       
        \label{fig:time_comparison}
    \end{subfigure}
    \hfill
    \begin{subfigure}[t]{0.58\textwidth}
        \centering
        \includegraphics[width=0.65\textwidth]{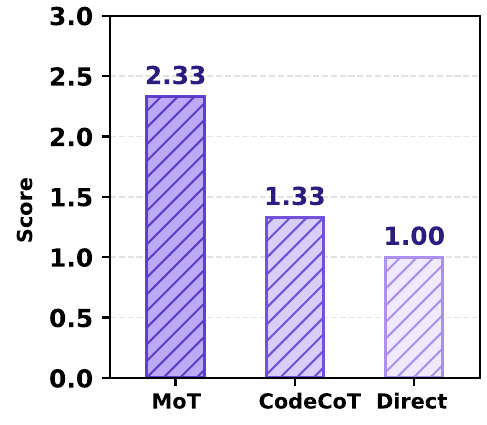}
        \caption{Evaluation on SWE-bench Lite}
        \label{fig:maintain}
    \end{subfigure}
    
    \caption{Performance and efficiency comparison across different prompting techniques}

\end{figure}

\textbf{\textit{1) Setup: }}
To answer this RQ, we conducted cost evaluation experiments on 164 HumanEval and 399 MBPP problems. Although variants such as HumanEval-ET/+ and MBPP-ET/+ exist, they differ only in the number of test cases. For accurate cost measurement, each problem was solved individually, and the computational cost was calculated based on token usage under different prompting techniques.

\textbf{\noindent\textit{2) Results: }}
We compare the cost of MoT with those of typical prompting techniques, as shown in Table \ref{tab:llm_cost}. 
Zero-shot and few-shot prompting techniques have notably lower computational costs compared to other prompting techniques. Specifically, zero-shot has the lowest In-token consumption, while few-shot achieves the lowest overall cost due to its lowest output-token consumption. Given the cost of an output token is typically much higher than that of an input token, few-shot has a lower cost than zero-shot. Although these two simple prompting techniques have low computational costs, they typically have inferior performance. The other results in Table \ref{tab:llm_cost} show that MoT demonstrates superior efficiency and lower computational cost compared to other typical prompting techniques.

Although MoT involves multiple generation steps, its intermediate outputs remain short and structured because the MLR graph follows a fixed three-level hierarchy and uses a predefined node-level reasoning schema. Moreover, the shortcut policy stops unnecessary expansion and compresses the reasoning for simple subtasks. 
Therefore, MoT can generate fewer output tokens than reasoning-oriented baselines such as CoT, Self-Planning, SCoT, and CodeCoT, which tend to generate longer free-form intermediate outputs.

As shown in Figure~\ref{fig:time_comparison}, MoT remains efficient, with an average generation time of 5 seconds.
Although few-shot is faster (3 seconds), 
it delivers much weaker generation performance. MoT is also faster than Self-Planning, SCoT, and CodeCoT, which take 6, 9, and 13 seconds, respectively. Unlike techniques such as CodeCoT, MoT does not require generating test cases, executing code, or iterative generation. Reducing unnecessary steps leads to a more efficient process, enabling MoT to achieve improved performance without compromising effectiveness.

\begin{tcolorbox}[
  enhanced,
  colback=grey,                
  colframe=teal!60!black,       
  boxrule=0.6pt,                
  arc=10pt,                     
  left=4mm,right=4mm,           
  top=2mm,bottom=2mm,
  drop shadow={black!40!white}, 
]
\textbf{\textit{Answer to RQ3:} }
\textit{Based on these results, we conclude that MoT offers an efficient solution to achieve better performance, which incurs significantly lower cost than other prompting methods except simple zero-shot and few-shot.
}

\end{tcolorbox}

\subsection{What Is the Impact of MLR Graph Depth on Code Generation Performance? (RQ4)}

\textbf{\textit{1) Setup: }}To analyse the effect of hierarchical depth in the MLR graph, we conduct an ablation study where we only vary the MLR graph depth and keep all other settings unchanged. We evaluate GPT-4o-mini on HumanEval and HumanEval+, measuring Pass@1 accuracy. To further examine whether the depth choice remains effective in a more realistic repository-level setting, we also evaluate different MLR graph depths on RAL-Bench using the same model.

\textbf{\textit{2) Results: }}Results are shown in Table \ref{tab:layers-ablation}. The results show that performance peaks at 3 layers, suggesting that this configuration provides a practical balance between abstraction and granularity in the studied settings. Concretely, on HumanEval, Pass@1 rises from 83.5 (1 layer) to 85.4 (2 layers) and reaches 92.1 at 3 layers. On HumanEval+, the same trend holds, where Pass@1 increases from 75.6 to 79.9 and reaches the best result of 83.5 at 3 layers. On RAL-Bench, the 3-layer configuration also achieves the best functional score of 25.8, compared with 22.1 for 1 layer, 22.2 for 2 layers, and 23.9 for 4 layers. In contrast, adding an extra layer hurts performance. The 4-layer configuration drops to 87.2 on HumanEval and 78.0 on HumanEval+, which are 4.9 and 5.5 points lower than the 3-layer configuration, respectively. A similar trend is observed on RAL-Bench, where the 4-layer configuration drops by 1.9 points compared with the 3-layer configuration. These results indicate that shallower graphs (1--2 layers) may lack sufficient structure to guide reasoning, because they often stop at coarse steps and fail to make intermediate implementation decisions explicit. However, deeper graphs (4 layers) may introduce unnecessary redundancy and fragmentation, potentially over-decomposing the task into overly fine-grained nodes and increasing coordination overhead during modular generation. 
Overall, the depth study shows that the 3-layer hierarchy provides the best empirical trade-off across the evaluated function-level and repository-level settings. 
MoT adopts a default 3-layer hierarchy as a practical choice for balancing decomposition sufficiency and generation overhead.

\begin{tcolorbox}[
  enhanced,
  colback=grey,                
  colframe=teal!60!black,       
  boxrule=0.6pt,                
  arc=10pt,                     
  left=4mm,right=4mm,           
  top=2mm,bottom=2mm,
  drop shadow={black!40!white}, 
]
\textbf{\textit{Answer to RQ4:} }
\textit{
Based on these results, we conclude that a 3-layer MLR hierarchy provides a strong practical default for balancing abstraction and granularity across the studied function-level and repository-level settings. Shallower hierarchies lack sufficient structured guidance, while deeper hierarchies introduce redundancy, fragmentation, and coordination overhead}. Therefore, MoT adopts 3 layers as the default depth for effective decomposition.

\end{tcolorbox}

\begin{table}[t]
\centering
\caption{Impact of MLR graph depth on function-level and repository-level code generation.}
\label{tab:layers-ablation}
\setlength{\tabcolsep}{4pt}
\renewcommand{\arraystretch}{1.08}
\small
\resizebox{0.8\linewidth}{!}{%
\begin{tabular}{l
S[table-format=2.1] S[table-format=+2.1]
S[table-format=2.1] S[table-format=+2.1]
S[table-format=2.1] S[table-format=+2.1]}
\toprule
\textbf{Graph Depth} &
\multicolumn{2}{c}{\textbf{HumanEval}} &
\multicolumn{2}{c}{\textbf{HumanEval+}} &
\multicolumn{2}{c}{\textbf{RAL-Bench}} \\
\cmidrule(lr){2-3}\cmidrule(lr){4-5}\cmidrule(lr){6-7}
& {\textbf{Pass@1}} & {\textbf{$\Delta$ vs Avg (\%)}} 
& {\textbf{Pass@1}} & {\textbf{$\Delta$ vs Avg (\%)}} 
& {\textbf{Fun.}} & {\textbf{$\Delta$ vs Avg (\%)}} \\
\midrule
1 layer   & 83.5 & -4.1 & 75.6 & -4.6 & 22.1 & -5.8 \\
2 layers  & 85.4 & -1.9 & 79.9 & +0.8 & 22.2 & -5.7 \\
3 layers  & {\bfseries 92.1} & +5.8 & {\bfseries 83.5} & +5.4 & {\bfseries 25.8} & +10.0 \\
4 layers  & 87.2 & +0.2 & 78.0 & -1.6 & 23.9 & +1.5 \\
\midrule
\rowcolor{cyan!8}
\textbf{Average} & 87.1 & 0.0 & 79.3 & 0.0 & 23.5 & 0.0 \\
\bottomrule
\end{tabular}%
}
\end{table}

\subsection{Is MoT Robust to Ambiguities and Noisy Variations in Prompts? (RQ5)}

\begin{table*}[t]
\centering
\caption{Comparative robustness under prompt perturbations on HumanEval using GPT-4o-mini. Delta values are computed as relative changes from the original-prompt setting to the perturbed-prompt setting. Smaller absolute changes indicate better robustness.}
\label{tab:robustness}
\resizebox{\textwidth}{!}{
\begin{tabular}{lccc|ccc|ccc}
\toprule
\multirow{2}{*}{\textbf{Method}} 
& \multicolumn{3}{c|}{\textbf{Synonym}} 
& \multicolumn{3}{c|}{\textbf{Shuffle}} 
& \multicolumn{3}{c}{\textbf{Typos}} \\
\cmidrule(lr){2-4} \cmidrule(lr){5-7} \cmidrule(lr){8-10}
& $\Delta$\textbf{Pass@1} & $\Delta$\textbf{BLEU} & $\Delta$\textbf{Edit}
& $\Delta$\textbf{Pass@1} & $\Delta$\textbf{BLEU} & $\Delta$\textbf{Edit}
& $\Delta$\textbf{Pass@1} & $\Delta$\textbf{BLEU} & $\Delta$\textbf{Edit} \\
\midrule
CoT           & -1.83\% & \textbf{+0.12\%} & -0.17\% & -50.00\% & \textbf{-1.01\%} & +0.86\% & +0.61\% & \textbf{+0.10\%} & -0.35\% \\
Self-Planning & +5.49\% & -0.51\% & +0.81\% & -47.56\% & -4.22\% & +7.08\% & -0.61\% & -0.55\% & +1.08\% \\
SCoT          & +1.83\% & -1.13\% & +0.70\% & -52.44\% & -3.53\% & +4.71\% & -2.44\% & -0.13\% & +0.30\% \\
CodeCoT       & \textbf{-1.22\%} & +0.33\% & \textbf{+0.10\%} & -54.88\% & -2.84\% & +3.59\% & -4.27\% & +0.11\% & -0.92\% \\
\rowcolor{cyan!8}
MoT           & \textbf{-1.22\%} & -0.55\% & -0.17\% & \textbf{-39.63\%} & -2.35\% & \textbf{+0.71\%} & \textbf{-0.34\%} & -0.12\% & \textbf{-0.29\%}  \\
\bottomrule
\end{tabular}
}
\end{table*}

\textbf{\textit{1) Setup: }} To study the robustness of different prompting methods to small variations or ambiguities in the problem description, we conducted a comparative evaluation on the HumanEval benchmark using GPT-4o-mini. We chose GPT-4o-mini because it is one of the main models in our study and provides a good balance between generation quality and evaluation cost for this comparative robustness analysis. We applied three types of controlled natural language perturbations to the input prompts:

\begin{itemize}
    \item \textbf{Synonym substitution} (e.g., ``calculate'' $\rightarrow$ ``compute'')
    \item \textbf{Word order shuffling} (within sentences)
    \item \textbf{Character-level typos} (e.g., ``return'' $\rightarrow$ ``retrun'')
\end{itemize}

We then generate code for both original and perturbed prompts using MoT and four representative baselines, including CoT, Self-Planning, SCoT, and CodeCoT. We evaluate robustness from three aspects: Pass@1, BLEU, and normalized edit distance. For each metric $m$, we first compute its value under the original-prompt setting, denoted as $m_{\mathrm{orig}}$, and under the perturbed-prompt setting, denoted as $m_{\mathrm{pert}}$. We then report the relative change as
\[
\Delta m = \frac{m_{\mathrm{pert}} - m_{\mathrm{orig}}}{m_{\mathrm{orig}}} \times 100\%.
\]
Therefore, $\Delta$Pass@1 measures the relative task-level performance change under perturbation, $\Delta$BLEU measures the relative change in BLEU score, and $\Delta$Edit measures the relative change in normalized edit distance. Smaller absolute values indicate that the method is less affected by the perturbation. Table~\ref{tab:robustness} summarizes the averaged comparative robustness results across all HumanEval tasks.

\textbf{\textit{2) Results: }}As shown in Table~\ref{tab:robustness}, synonym substitution causes only small changes for most methods, suggesting that semantically equivalent wording changes are generally easy to handle. Word order shuffling is the most disruptive perturbation, but MoT shows the smallest Pass@1 drop (-39.63\%) and the smallest absolute $\Delta$Edit among all methods. Under character-level typos, MoT also shows the smallest Pass@1 change (-0.34\%) and a small edit-distance change (-0.29\%). Overall, MoT is more stable than representative prompting baselines under the more disruptive perturbations.

\begin{tcolorbox}[
  enhanced,
  colback=grey,                
  colframe=teal!60!black,       
  boxrule=0.6pt,                
  arc=10pt,                     
  left=4mm,right=4mm,           
  top=2mm,bottom=2mm,
  drop shadow={black!40!white}, 
]
\textbf{\textit{Answer to RQ5:} }
\textit{
MoT shows competitive robustness under noisy and ambiguous prompt perturbations. It achieves smaller performance changes in several settings, while word-order shuffling remains a challenging case.}

\end{tcolorbox}

\section{Discussion}

\subsection{How Does MoT Perform on Other Real-World Software Development Tasks?}

The RQ1 evaluation has already included LiveCodeBench and RAL-Bench to assess MoT on recent code generation tasks and repository-level generation tasks. 
In this section, we further examine MoT on real-world repository bug fixing. 
To this end, we evaluate MoT on the official \textbf{SWE-bench Lite} set, which contains 300 real-world software issue tasks, using GPT-4o-mini. 
This experiment provides preliminary evidence on whether MoT can help with repository-level bug fixing. For each task, all methods receive the same issue description and the same repository context, and are evaluated under the same patch-generation and test-execution protocol. This setting keeps the comparison focused on the prompting strategy under a shared input context. We compare MoT with two representative baselines. \textit{Direct} serves as a one-pass generation baseline without explicit decomposition or structured reasoning. \textit{CodeCoT} is selected because it is one of the strongest prompting baselines in RQ1 and incorporates self-examination during code generation. This comparison allows us to examine whether MoT brings benefits beyond direct generation and reasoning-based prompting.

Figure~\ref{fig:maintain} compares Direct, CodeCoT, and MoT on SWE-bench Lite. Overall, MoT achieves competitive performance. These results suggest that MoT can be applied to repository editing tasks beyond standalone code generation. 
More comprehensive evaluation on real-world software engineering benchmarks, especially with stronger retrieval, localization, and feedback-driven repair mechanisms, remains future work.

\subsection{How Does MoT Compare with Agent-Based Methods?}

To complement the comparisons with prompting-only baselines in RQ1, we further compare MoT with representative agent-based repository-level code generation methods. 
We present this comparison separately from RQ1 because agent-based methods differ from prompting-only baselines in both inference procedure and computational cost.
While RQ1 focuses on prompting baselines evaluated under the same experimental protocol, agent-based methods often involve multi-step coordination, search, or iterative refinement \citep{lin2025se}.
This comparison helps clarify how MoT performs relative to agent-based repository-level methods.

We evaluate MoT on RAL-Bench~\citep{pan2026ral} and NL2Repo-Bench~\citep{ding2025nl2repo} for this separate comparison. RAL-Bench is also used in the main evaluation, while NL2Repo-Bench is included here because its difficulty-stratified repository-level tasks are suitable for comparing MoT with agent-based methods. Together, these two benchmarks provide complementary views of repository-level generation. 
For NL2Repo-Bench, which is used only in this separate discussion comparison, we report \textbf{Overall}, \textbf{Easy}, \textbf{Medium}, and \textbf{Hard} scores following the original benchmark protocol~\citep{ding2025nl2repo}. Each score is computed as the average pytest pass rate of generated repositories.
We compare MoT with four representative baselines. \textit{Direct} is a simple one-pass generation baseline without explicit decomposition. \textit{SE-Agent} is a software-engineering-oriented agent baseline for repository-level tasks. \textit{AlphaEvolve} \citep{novikov2025alphaevolve} and \textit{CSE} \citep{hu2026controlled} are repository-level baselines that move beyond simple direct prompting. 
We use GPT-5 as the shared backbone for all methods so that the comparison is conducted under the same model setting and focuses on
the generation strategy of each method.

\begin{table*}[t]
\centering
\caption{Comparison with agent-based baselines on repository-level generation benchmarks.}
\label{tab:agent_compare}
\resizebox{0.7\textwidth}{!}{
\begin{tabular}{lccccc}
\toprule
\textbf{Method} 
& \multicolumn{4}{c}{\textbf{NL2Repo-Bench (GPT-5)}} 
& \multicolumn{1}{c}{\textbf{RAL-Bench (GPT-5)}} \\
\cmidrule(lr){2-5}
\cmidrule(lr){6-6}
& \textbf{Overall} & \textbf{Easy} & \textbf{Medium} & \textbf{Hard} & \textbf{Fun.} \\
\midrule
Direct & 21.7 & 38.4 & 20.7 & 9.6 & 38.5 \\
SE-Agent & 25.4 & 44.5 & 23.6 & 12.6 & 44.7 \\
AlphaEvolve & 25.1 & 41.8 & 25.1 & 11.7 & 43.7 \\
CSE & 19.7 & 41.1 & 15.8 & 7.9 & 34.3 \\
\rowcolor{cyan!8}
MoT & \textbf{29.7} & 34.8 & \textbf{36.8} & \textbf{20.7} & \textbf{46.2} \\
\bottomrule
\end{tabular}
}
\end{table*}

As shown in Table~\ref{tab:agent_compare}, MoT achieves the best results on both benchmarks. Its advantage is especially clear on the medium and hard subsets of NL2Repo-Bench. It also obtains the highest functional score on RAL-Bench. A likely reason is that MoT introduces a generation-aware hierarchical plan that directly connects task decomposition with downstream code generation, while avoiding the extra coordination or search complexity of agent-based methods. 
These results suggest that MoT is effective for repository-level code generation tasks.

\subsection{How Does MoT Compare with Structured Reasoning Methods?}

We present this comparison separately from RQ1 because these methods differ markedly from standard prompting baselines in both inference procedure and cost.
While RQ1 focuses on standard prompting baselines, this subsection examines stronger structured reasoning methods, including ToT, GoT, Parsel, CodeChain, and CodeTree. We evaluate these methods on HumanEval and RAL-Bench, which provide complementary views of code generation performance. HumanEval measures function-level code generation under a standard benchmark setting, while RAL-Bench evaluates functional effectiveness on more realistic repository-level tasks. We use GPT-4o-mini as the shared backbone.
All compared methods are reproduced based on their original papers or released implementations.
We therefore report both effectiveness and full-benchmark cost to provide a more faithful comparison.

\begin{table*}[t]
\centering
\caption{Performance and full-benchmark cost comparison with structured reasoning and refinement methods on HumanEval and RAL-Bench using GPT-4o-mini.}
\label{tab:structured_compare}
\resizebox{0.9\textwidth}{!}{
\begin{tabular}{lcc|cc|cc}
\toprule
\multirow{2}{*}{\textbf{Method}} 
& \multicolumn{2}{c|}{\textbf{Performance}} 
& \multicolumn{2}{c|}{\textbf{HumanEval Cost}} 
& \multicolumn{2}{c}{\textbf{RAL-Bench Cost}} \\
\cmidrule(lr){2-3} \cmidrule(lr){4-5} \cmidrule(lr){6-7}
& \textbf{HumanEval Pass@1} & \textbf{RAL-Bench Fun.}
& \textbf{Token} & \textbf{Time (s)}
& \textbf{Token} & \textbf{Time (s)} \\
\midrule
ToT       & 84.1 & 8.5  & 1817.3 & 15.58 & 4949.0 & 36.81 \\
GoT       & 90.2 & 16.6 & 2712.3 & 20.04 & 4546.1 & 34.99 \\
Parsel    & 71.8 & 14.0 & 2784.7 & 19.95 & 5489.4 & 32.14 \\
CodeChain & 90.9 & 15.3 & 4323.0 & 25.43 & 8389.9 & 45.93 \\
CodeTree  & 84.8 & 13.3 & 3503.6 & 21.18 & 9722.1 & 52.06 \\
\rowcolor{cyan!8}
MoT       & \textbf{92.1} & \textbf{25.8 } & 783.7 & 5.12 & 3431.6 & 31.69 \\
\bottomrule
\end{tabular}
}
\end{table*}

As shown in Table~\ref{tab:structured_compare}, MoT remains competitive with closely related structured reasoning methods on both HumanEval and RAL-Bench, while also maintaining favorable efficiency. This suggests that MoT can achieve strong performance without relying on the heavier search or refinement procedures used by some alternative methods.

\subsection{Does MoT Lead to Over-Modularization?}

\textbf{Over-modularization is the core risk.} A common concern of modular prompting is over-modularization, where the generated solution introduces unnecessary functions or modules. From a software engineering perspective, such over-modularization can increase complexity, reduce readability, and add avoidable indirection. This risk is most pronounced if modular reasoning units are naively mapped to isolated functions.

\textbf{MoT reduces over-modularization risk.} MoT mitigates this risk in two ways. First, it fixes decomposition to three levels, which bounds granularity and prevents unbounded splitting into micro-tasks. Second, MoT does not require each node to become a separate function: reusable nodes can be implemented as helpers, while simple nodes can be inlined. For simple nodes, we also keep the reasoning concise to avoid pushing the model toward over-modularization. 
This aligns with our depth study: decomposition helps up to a point, whereas over-modularization can introduce redundancy, coordination overhead, and unnecessary function-call overhead.
Overall, MoT should be viewed as a planning framework rather than a strict rule for function boundaries. MoT is not designed as a backtracking-based repair framework, but as a planning scaffold for improving first-pass generation. Still, the explicit MLR graph improves traceability by exposing node-level decomposition and reasoning, which can make failures easier to localize than in a monolithic generation process. In this sense, MoT is complementary to feedback-driven repair methods rather than a replacement for them. Future work could make the modularization more adaptive by deciding more directly when to inline code and when to create a helper function.

\section{Threats to Validity}

The first potential threat relates to the stochasticity of LLMs, 
which could affect the measured accuracy and stability of the results. To mitigate this threat, we repeat experiments across benchmarks and report the averaged results. This reduces the influence of random fluctuations and improves the robustness of our empirical observations.

The second potential threat concerns the generalizability of MoT across programming languages. Our current evaluation primarily focuses on Python, while programming languages differ in syntax, abstraction mechanisms, and library ecosystems, which may affect the effectiveness of MoT. 
We choose Python because it is widely used and representative in code-generation evaluation. In future work, we plan to further validate MoT on a broader range of programming languages.

The third potential threat concerns the generalizability of our evaluation to broader software engineering tasks. Evidence from function-level benchmarks and repository-level generation alone may be insufficient to support generality across diverse maintenance scenarios. To mitigate this threat, we extend our evaluation to repository-level generation settings, including RAL-Bench and a separate comparison on NL2Repo-Bench, and include a preliminary SWE-bench Lite study on repository-level bug fixing. However, our current maintenance-oriented evaluation is still limited to bug fixing and does not fully cover other tasks such as feature enhancement. More comprehensive evaluation on broader software engineering tasks, such as FEA-Bench~\citep{li2025fea}, remains future work.

The fourth potential threat concerns the suitability of the fixed three-level MLR hierarchy. The most appropriate hierarchy depth may vary across tasks and settings. To mitigate this threat, we conduct a dedicated depth analysis and compare multiple hierarchy depths in the current evaluation setting. Our results show that a three-level hierarchy works well for the studied tasks. Exploring the optimal hierarchy depth across diverse settings remains future work.

\section{Conclusion}

In this paper, we propose a novel prompting technique MoT to improve the code generation performance of LLMs. Different from the existing prompting techniques, MoT 
incorporates the modularization principles into the reasoning process. 
It introduces a novel Multi-Level Reasoning (MLR) graph to enhance modular understanding and ensure alignment between reasoning and the generated code by hierarchically organizing reasoning steps with its modular structure.
We have conducted an extensive study with two advanced LLMs on eight benchmarks. 
We further provide additional analyses on NL2Repo-Bench and SWE-bench Lite.
The results confirm the superiority of MoT over state-of-the-art prompting techniques. In future work, we will extend MoT to multi-language and real-world software tasks, and explore adaptive decomposition strategies that better balance structure and over-modularization.

\section{Data Availability}

To ensure the reproducibility of our results and to provide transparency in our research, we have made all related scripts and data publicly available. All resources (our source code, experimental data, and concrete examples of prompts and generated code) can be accessed as part of our artifact, which is available at \href{https://github.com/Wwstarry/Modularization_of_Thought}
     {\textbf{\textcolor{blue}{\nolinkurl{https://github.com/Wwstarry/Modularization_of_Thought}}}}.




\bibliographystyle{ACM-Reference-Format}
\bibliography{references}

\end{document}


\section*{Appendix}

\subsection{Appendix A Model Selection Justification}

To ensure practical and diverse model evaluation, we selected \textbf{GPT-4o-mini} and \textbf{DeepSeek-R1} (671B) because of their exceptional \textit{Pass@1} performance and cost-efficiency. As shown in Table \ref{tab:model-comparison}, these two models achieve top-tier accuracy on \textbf{HumanEval} and \textbf{HumanEval+} benchmarks while maintaining minimal API usage cost—especially GPT-4o-mini. Together, they represent two distinct paradigms: general-purpose LLMs and reasoning-specialized LLMs. Their favorable performance-to-cost ratio makes them ideal for scalable experimentation across a wide range of code generation tasks.

\begin{table}[htbp]
\centering
\small

\caption{Pass@1 score (\%) and API cost comparison of models.}

\setlength{\tabcolsep}{2pt}
\begin{tabular}{lccccc}

\toprule
\textbf{Model} & 
\makecell{\textbf{HumanEval} \\(Pass@1)} & 
\makecell{\textbf{HumanEval+} \\(Pass@1)} & 
\makecell{\textbf{Input} \\ \textbf{Cost} \\ (\$/1M)} & 
\makecell{\textbf{Output} \\ \textbf{Cost} \\ (\$/1M)} \\
\midrule
DeepSeek-R1        & \textbf{93.3}  & \textbf{93.1}  & --     & --     \\
GPT-4 Turbo        & 90.2  & 86.6  & 10.00  & 30.00  \\
DeepSeek-V3        & 91.5  & 86.6  & --     & --     \\
Claude 3.5 Sonnet  & 88.4  & 81.7  & 3.00   & 15.00  \\
GPT-4o-mini        & 88.4  & 81.1  & \textbf{0.15} & \textbf{0.60} \\
\bottomrule
\end{tabular}

\label{tab:model-comparison}
\end{table}

\subsection{Appendix B Details of the Baselines}

To evaluate MoT, we consider six typical prompting techniques for comparisons:

\begin{itemize}
    \item \textbf{Zero-shot prompting} \citep{chen2021evaluating} is a method of generating code without utilizing any code examples. The model generates code based solely on problem descriptions.
    \item \textbf{Few-shot prompting} \citep{chen2021evaluating} allows LLMs to choose a few examples for understanding the relationship between problem and code. In our experiments, we adopt the 2-shot setting. 
    \item \textbf{CoT prompting} \citep{wei2022chain} addresses complex code problems through step-by-step reasoning. When the model approaches a code problem, it first produces a sequence of intermediate steps, enhancing the logical coherence and correctness of the generated code.
    \item \textbf{Self-planning prompting} \citep{jiang2024self} enables the model to formulate a step-by-step plan prior to code generation. The model creates a comprehensive plan for code generation based on the problem description, which it then executes incrementally.
    \item \textbf{SCoT prompting} \citep{li2025structured} builds upon CoT prompting by leveraging program structures (i.e., sequence, branch, and loop structures) to generate intermediate reasoning steps. 
    \item \textbf{CodeCoT prompting} \citep{huang2023codecot} integrates CoT with self-examination mechanisms. The model improves the accuracy of generated code by producing logically coherent initial code and corresponding test cases, verifying code execution, and iteratively resolving errors as they are identified.
\end{itemize}

\subsection{Appendix C The Detailed Explanation About Metrics}

\textit{\underline{Pass@1}} measures the functional correctness of the generated code.
This metric is used to evaluate the performance of generated code in test cases. Given a programming problem, LLM generates one code instance. The problem is considered solved only if the instance passes all test cases. Pass@1 is the percentage of solved problems out of the total number of problems. The formula is as follows:

\begin{equation}
\text{Pass@1} := \mathbb{E}_{\text{Problems}} \left[ 1 - \frac{n-c}{n} \right]
\end{equation}

\textit{\underline{AvgPassRatio}} measures the correctness of the generated code based on its performance across evaluation test cases. Pass@1 focuses on whether the generated code is completely correct in the test case, so we introduce AvgPassRatio to complement it. AvgPassRatio (APR) is calculated by determining the ratio of passed evaluation test cases to the total number of evaluation test cases for each problem and then averaging this ratio across all problems. Larger AvgPassRatio values indicate better code generation performance. Note that in our experiments, APR was not calculated for the MBPP+ dataset as it lacks complete test cases.

\subsection{Appendix D Is MoT Robust to Ambiguities and Noisy Variations in Prompts?}

To address Question 3 regarding the robustness of MoT to small variations or ambiguities in the problem description, we conducted a dedicated study using the HumanEval benchmark. We applied three types of controlled natural language perturbations to the input prompts:

\begin{itemize}
    \item \textbf{Synonym substitution} (e.g., ``calculate'' $\rightarrow$ ``compute'')
    \item \textbf{Word order shuffling} (within sentences)
    \item \textbf{Character-level typos} (e.g., ``return'' $\rightarrow$ ``retrun'')
\end{itemize}

We then generated code for both original and perturbed prompts using MoT and measured the similarity of outputs via BLEU score and normalized edit distance. Table~\ref{tab:robustness} summarizes the averaged results overall HumanEval tasks.

\begin{table}[h]
\centering

\begin{tabular}{lcc}
\toprule
\textbf{Perturbation Type} & \textbf{BLEU $\uparrow$} & \textbf{Edit Distance $\downarrow$} \\
\midrule
Synonym & 0.9173 & 0.0277 \\
Shuffle & 0.5701 & 0.1450 \\
Typos   & 0.7389 & 0.0929 \\
\bottomrule
\end{tabular}
\caption{Robustness of MoT under prompt perturbations.}
\label{tab:robustness}
\end{table}

These results indicate that MoT is highly robust to semantically equivalent transformations and exhibits stable behavior even under noisy inputs. Although this study was conducted only on HumanEval, it suggests that the MLR Graph structure effectively preserves task semantics and helps maintain code generation quality despite input ambiguities. 

\subsection{Appendix E Why is MoT Effective?}

MoT constructs an MLR graph to delineate different levels of tasks, such as high-level, intermediate-level, and detailed-level tasks. This modular approach provides the LLM with a clear hierarchical framework during the reasoning process. By explicitly defining task boundaries and dependencies, the MLR graph guides the model to systematically address each module, thereby mitigating ambiguity and promoting logical clarity. 
Unlike traditional linear prompting methods that rely on sequential reasoning without explicit structure, MoT significantly reduces the potential for context confusion or unclear dependencies between steps. Moreover, by decomposing complex problems into modular sub-tasks, MoT reduces cognitive complexity and enables the model to more accurately and efficiently manage each sub-task. This modular reasoning not only improves the overall coherence of the generated code but also enhances maintainability and scalability, making it easier to debug, test, and extend the generated solutions. Consequently, MoT demonstrates substantial improvements in code generation quality when handling sophisticated programming problems.

\begin{figure}[h]
    \centering
    \includegraphics[width=0.5\textwidth]{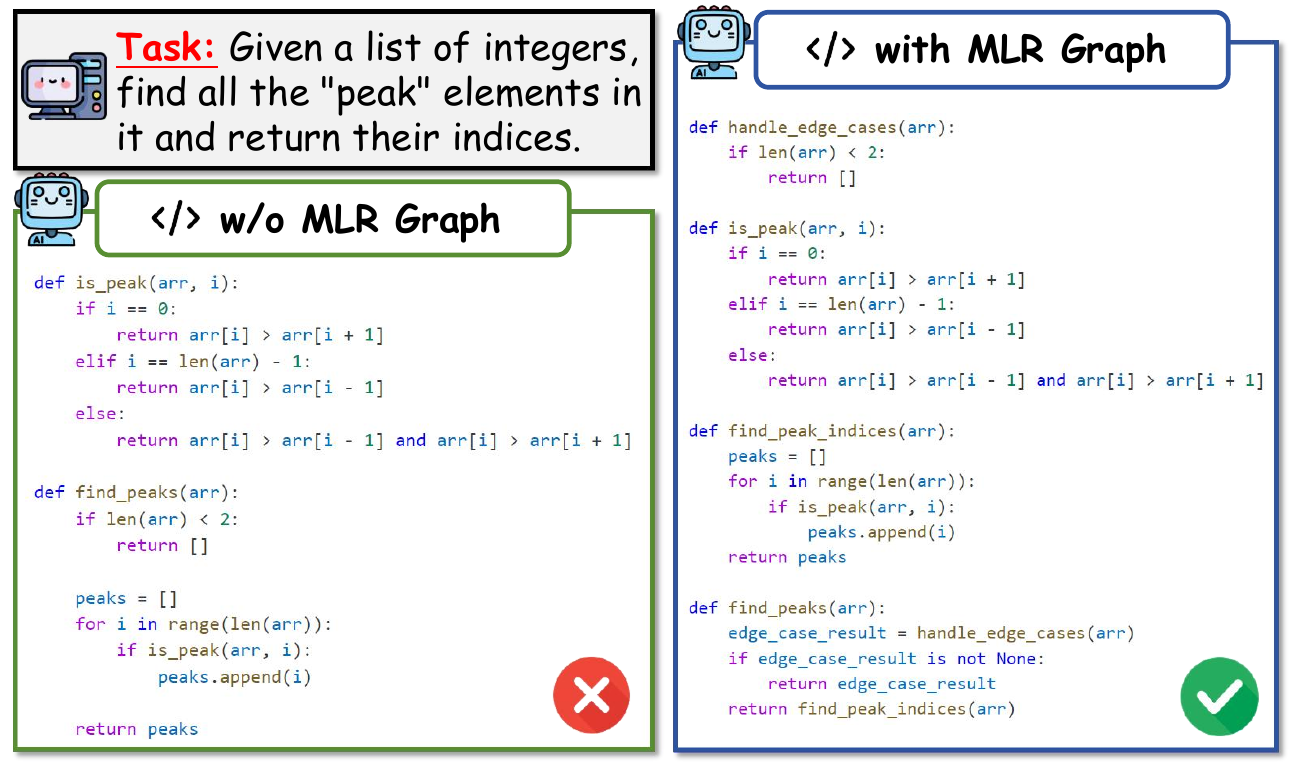}
    \caption{An example for the effectiveness of MLR Graph}
    \label{fig:MLR_example_effective}
\end{figure}

To illustrate the effectiveness of the MLR Graph in modular code generation, we analyze the specific programming task shown in Figure \ref{fig:MLR_example_effective}. The task requires identifying all "peak" elements in a given list of integers and returning their indices. The left side of the figure displays the code generated without MLR graph guidance, while the right side presents the code generated with MLR graph guidance. From the figure, it is evident that the right-side implementation, guided by the MLR graph, follows a well-structured and modular approach. It first handles edge cases (when the list length is less than 2), then separately defines two key logical components: one for determining whether an element is a peak (\textit{is\_peak}) and another for finding the indices of all peak elements (\textit{find\_peak\_indices}). Finally, the main function \textit{find\_peaks} integrates these modular components to complete the task. This modularization improves both the readability and maintainability of the generated code. The code on the left, which does not utilize MLR graph, lacks explicit handling for boundary cases (e.g., when the array length is less than 2, an index out-of-bounds error may occur). Additionally, its reasoning is monolithic, reducing both readability and robustness. Consequently, this implementation is prone to logical flaws and incomplete boundary case handling, making it more susceptible to unexpected errors. This example demonstrates that the MLR graph effectively guides LLMs in modular task decomposition and modular code generation. By leveraging this modular approach, MLR Graph enhances code quality and improves readability.

\subsection{Appendix F MoT for Code Maintainability}

MoT is aligned with good practices in software engineering, such as information hiding and modular decomposition, which have been shown to enhance system flexibility, maintainability, and scalability \citep{parnas1978buzzword, parnas1972criteria}. By incorporating these principles into the reasoning process of LLMs, 
MoT employs a modular reasoning approach to decompose complex code generation tasks into multiple independent yet interrelated modules, enabling each reasoning step to be optimized, iterated, and adjusted independently. Unlike CoT, which follows a monolithic reasoning process where all steps are highly coupled — making local error correction challenging. MoT allows for localized adjustments to reasoning steps, thereby reducing error propagation.

\begin{figure}[h]
    \centering
    \includegraphics[width=0.5\textwidth]{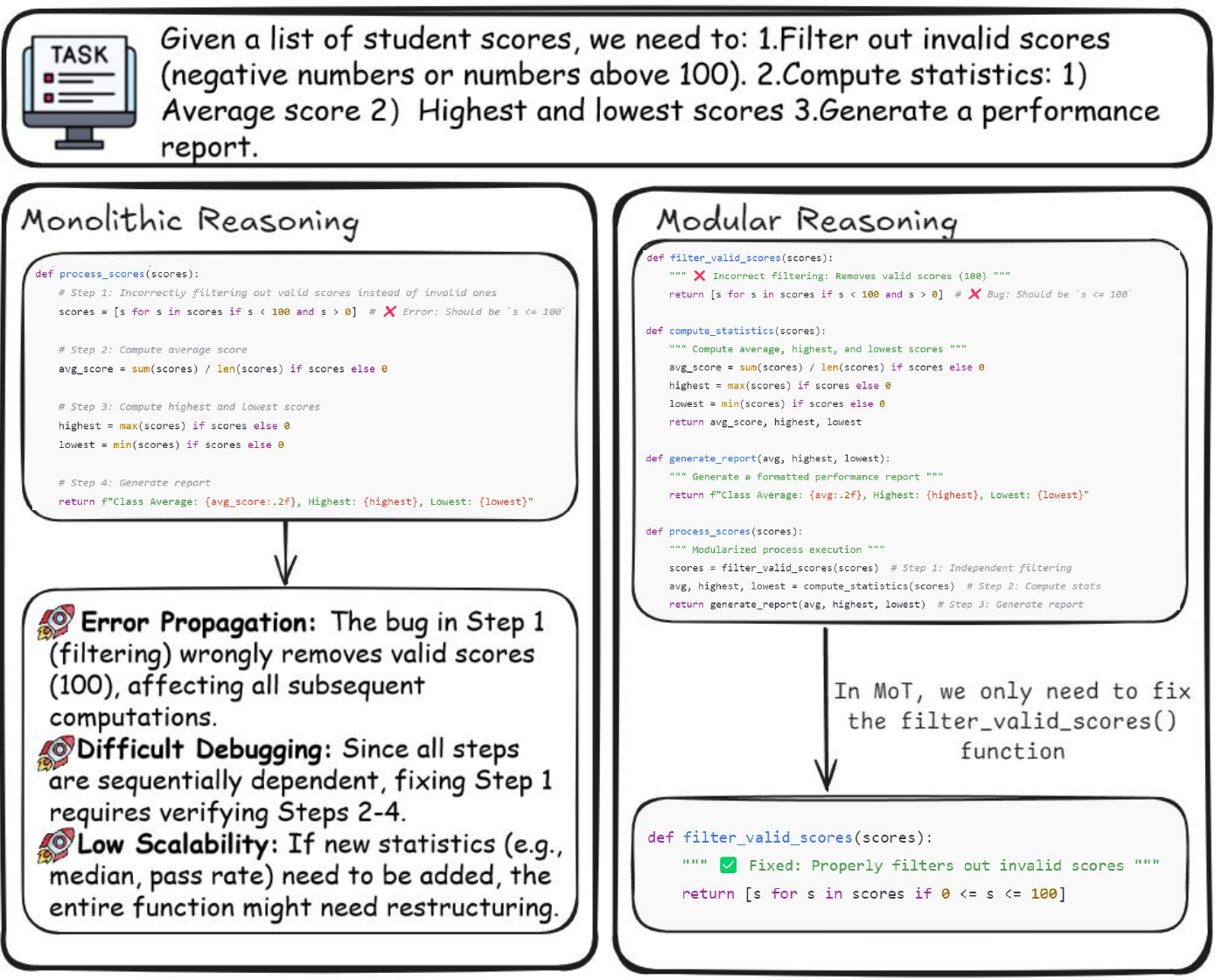}
    \caption{An example for MoT for Code Maintainability}

    \label{fig:effective}
\end{figure}

We provide an illustrative example in Figure \ref{fig:effective} to demonstrate the advantages of Modularized Reasoning compared to traditional monolithic reasoning. In traditional monolithic reasoning approaches (left side of Figure \ref{fig:effective}), all logic is encapsulated within a single module or function (e.g., the function \textit{process\_scores()}). This structure inherently causes any error within an intermediate step, such as incorrectly filtering valid scores, to propagate through subsequent computations. As a result, identifying the origin of such errors becomes cumbersome, requiring verification of multiple sequentially dependent steps. Moreover, extending such monolithic functions to accommodate new statistics or additional logic becomes challenging, reducing scalability and maintainability.

In contrast, as illustrated on the right side of Figure \ref{fig:effective}, the Modular Reasoning (MoT) decomposes the task into multiple independent yet interrelated modules. Each module, such as  \textit{filter\_valid\_scores()}, \textit{compute\_statistics()}, and \textit{generate\_report()}, is clearly defined and encapsulated with specific functionality. Consequently, debugging and correcting errors—like the improper filtering step—become simpler and more targeted, requiring modification of only the relevant module (\textit{filter\_valid\_scores()}). Furthermore, adding new functionalities or statistics can be achieved by introducing or updating individual modules without necessitating substantial restructuring of the overall code. Thus, modular reasoning significantly enhances the efficiency, scalability, and reliability of the generated code.

\subsection{Appendix G Threats to Validity}
The main limitations of our work lie in the following two aspects: Firstly, due to the randomness involved in LLMs, different outputs may be provided with each generation. As a result, even with identical input, MoT may yield different outputs, which can impact the accuracy and consistency of the final code generation. 
To solve this issue, we conduct extensive experiments to minimize influences caused by randomness involved in LLMs, such as performing repeated experiments on various benchmarks. This approach ensures consistent results in most cases and helps reduce unusual fluctuations in individual tests by evaluating averages.

Secondly, our current analysis primarily focuses on Python code generation, while the characteristics of other programming languages may not be fully explored. Different programming languages have distinct syntax and structures, which may result in varying effectiveness and performance of MoT. To address this limitation, we selected Python, a widely used and representative language, for the experiments. 
However, the modularization concept of MoT is applicable to most programming languages. In our future work, we plan to conduct more comprehensive evaluations across a broader range of programming tasks such as system programming and web development to further explore the adaptability and performance of MoT. 

\subsection{Appendix H Computer Infrastructure}

All experiments are conducted on three NVIDIA RTX A800 GPUs with 80GB of memory, and the LLMs are accessed through official APIs. The operating system is Ubuntu 22.04 LTS. Key libraries include Python 3.10 and CUDA 12.1.

\subsection{Appendix I Original Prompt Templates} 

We include below the exact prompt templates used for each baseline prompting technique in our experiments.

\begin{figure}[H]
\centering
\begin{lstlisting}[style=promptstyle,caption={Chain-of-Thought (CoT) Prompt}]
Please solve the problem step by step. Please provide the chain of thought briefly and give the final code.

Here is the problem. Please only provide its CoT and the final code: Problem Description
\end{lstlisting}
\end{figure}

\begin{figure}[H]
\centering
\begin{lstlisting}[style=promptstyle,caption={Self-Planning Prompt}]
Output Format:
def function_name(args):
    '''
    <Problem Description>
    Step-by-Step Plan:
    1. Step 1 description
    2. Step 2 description
    3. Step 3 description
    '''
    Implementation based on the plan

Here is an output example:
def minSubArraySum(nums):
    '''
    Given an array of integers nums, find the minimum sum of any
    non-empty sub-array of nums.
    Example:
    minSubArraySum([2, 3, 4, 1, 2, 4]) == 1  
    minSubArraySum([-1, -2, -3]) == -6  
    Step-by-Step Plan:
    1. Create a function to calculate the sum of a sub-array.
    2. Loop through the input list, calculating the sum of each sub-array.
    3. Update and return the minimum sum.
    '''
    # Helper function to calculate the sum of a sub-array
    def subArraySum(nums):
        sum = 0
        for i in nums:
            sum += i
        return sum
    # Implementation starts here
    min_sum = subArraySum(nums)  # Initialize with full array sum
    for i in range(len(nums)):
        for j in range(i + 1, len(nums) + 1):
            current_sum = subArraySum(nums[i:j])
            if current_sum < min_sum:
                min_sum = current_sum
    return min_sum

Here is the problem. Please only provide its Plan and the final code: <Problem Description>
\end{lstlisting}
\end{figure}

\newpage

\begin{figure*}[t]
\centering
\begin{lstlisting}[style=promptstyle,caption={Few-shot Prompt}]
Here are some examples:

Problem:
from typing import List

def has_close_elements(numbers: List[float], threshold: float) -> bool:
    """ Check if in given list of numbers, are any two numbers closer to each other than
    given threshold.
    >>> has_close_elements([1.0, 2.0, 3.0], 0.5)
    False
    >>> has_close_elements([1.0, 2.8, 3.0, 4.0, 5.0, 2.0], 0.3)
    True
    """

Output:
for idx, elem in enumerate(numbers):
    for idx2, elem2 in enumerate(numbers):
        if idx != idx2:
            distance = abs(elem - elem2)
            if distance < threshold:
                return True
return False

---

Problem:
from typing import List

def separate_paren_groups(paren_string: str) -> List[str]:
    """ Input to this function is a string containing multiple groups of nested parentheses. Your goal is to
    separate those group into separate strings and return the list of those.
    Separate groups are balanced (each open brace is properly closed) and not nested within each other
    Ignore any spaces in the input string.
    >>> separate_paren_groups('( ) (( )) (( )( ))')
    ['()', '(())', '(()())']
    """

Output:
result = []
current_string = []
current_depth = 0

for c in paren_string:
    if c == '(':
        current_depth += 1
        current_string.append(c)
    elif c == ')':
        current_depth -= 1
        current_string.append(c)

        if current_depth == 0:
            result.append(''.join(current_string))
            current_string.clear()

return result

Here is the problem. Please only provide the final code: Problem Description
\end{lstlisting}
\end{figure*}

\newpage

\begin{figure}[t]
\centering
\begin{lstlisting}[style=promptstyle,caption={SCoT Prompt}]
### Here is an example:
Please understand the requirement and write a rough solving process. It starts with an input-output structure.
You should use three basic structures to build the solving process, including sequences, branches, and loops.
The necessary details should be written in natural language.
Problem:
Write a python function to find the first repeated character in a given string.
SCoT:
Input: str: a string  
Output: ch: a repeated character in str  
1: for each character ch in str:  
2: if ch appears more than once in str:  
3: return ch  
4: return None

Please understand the requirement and write a rough solving process. It starts with an input-output structure. You should use three basic structures to build the solving process, including sequences, branches, and loops. The necessary details should be written in natural language.

Here is the problem. Please only provide its SCoT and the final code: <Problem Description>
\end{lstlisting}
\end{figure}

\FloatBarrier

\subsection{Appendix J Illustrative Example: Problem, Prompts, MLR Graph, and Generated Code}

This section presents an \textbf{end-to-end example} demonstrating how the \textbf{Multi-Level Reasoning (MLR) Graph} facilitates modular code generation.  
We provide: Problem, Prompts, MLR Graph, and Generated Code.

\begin{figure}[H]
\centering
\begin{minipage}{0.5\textwidth}
\begin{lstlisting}[style=promptstyle]
You are a code reasoning assistant. Your task is to analyze the given programming problem and generate a modular reasoning graph (Multi-Level Reasoning Graph, MLR Graph) to guide the code generation process. Provide your reasoning in the following hierarchical textual format clearly:

Format:
H1 [High-Level]: Solve the problem: {Problem description}
  Reasoning: Break the problem into major tasks: {High-level task 1} and {High-level task 2}.
  ├── H1.1 [High-Level]: {Subtask 1 of High-Level}
  │       Reasoning: {Reasoning for subtask 1}
  │       ├── I1.1 [Intermediate-Level]: {Intermediate-level task 1 for subtask 1}
  │       │         Reasoning: {Reasoning for intermediate-level task 1}
  │       └── I1.2 [Intermediate-Level]: {Intermediate-level task 2 for subtask 1}
  │                 Reasoning: {Reasoning for intermediate-level task 2}
  ├── H1.2 [High-Level]: {Subtask 2 of High-Level}
  │       Reasoning: {Reasoning for subtask 2}
  │       ├── I2.1 [Intermediate-Level]: {Intermediate-level task 1 for subtask 2}
  │       │         Reasoning: {Reasoning for intermediate-level task 1}
  │       │         └── D2.1 [Detailed-Level]: {Detailed implementation details or pseudo-code}
\end{lstlisting}
\captionsetup{type=figure}
\caption{Prompt Template for MLR Graph Generation}
\end{minipage}
\end{figure}

\begin{figure}[H]
\centering
\begin{minipage}{0.5\textwidth}
\begin{lstlisting}[style=promptstyle]
You are a code generation assistant. Your task is to generate modular code based on the given modular reasoning (MLR graph). You must only output the generated code.
Modular Reasoning (MLR graph): {MLR_graph}. 
Output: Provide only the complete code corresponding to the given modular reasoning. If possible, organize the code into multiple modular functions.
\end{lstlisting}
\captionsetup{type=figure}
\caption{Prompt Template for Code Generation}
\end{minipage}
\end{figure}

\begin{figure}[H]
\centering
\includegraphics[width=0.5\textwidth]{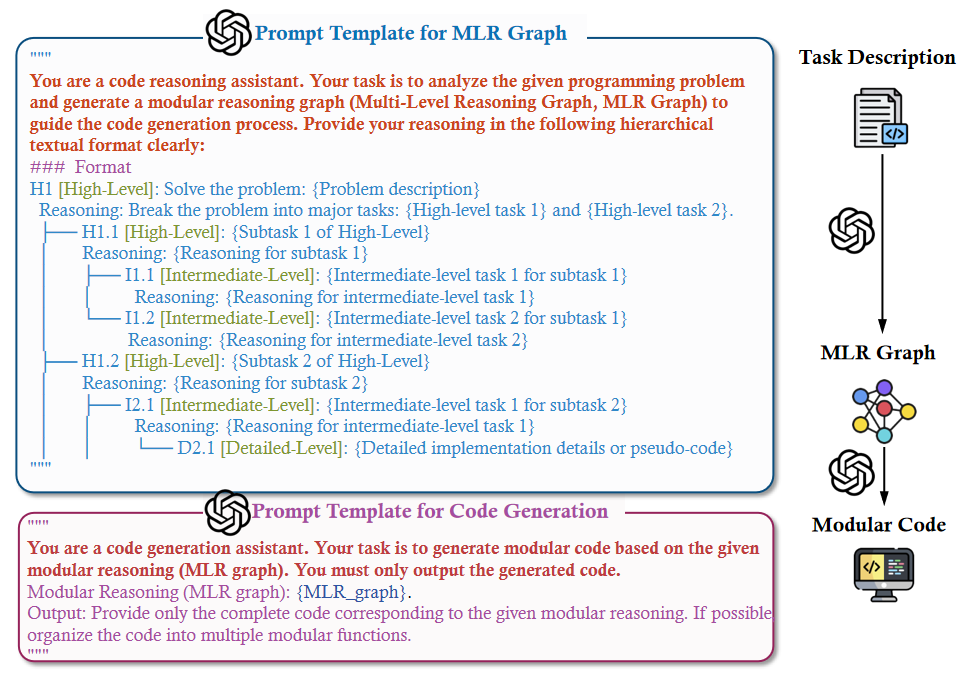}
\caption{Visualization of the prompting process for MLR Graph and code generation.}
\end{figure}

\textbf{Task:} You are given a list of lists (an array of arrays) and an integer \textbf{K}.  
The goal is to \textbf{find the largest sum among the sublists} and then \textbf{divide this sum by K}.

\begin{figure}[H]
\centering
\includegraphics[width=0.5\textwidth]{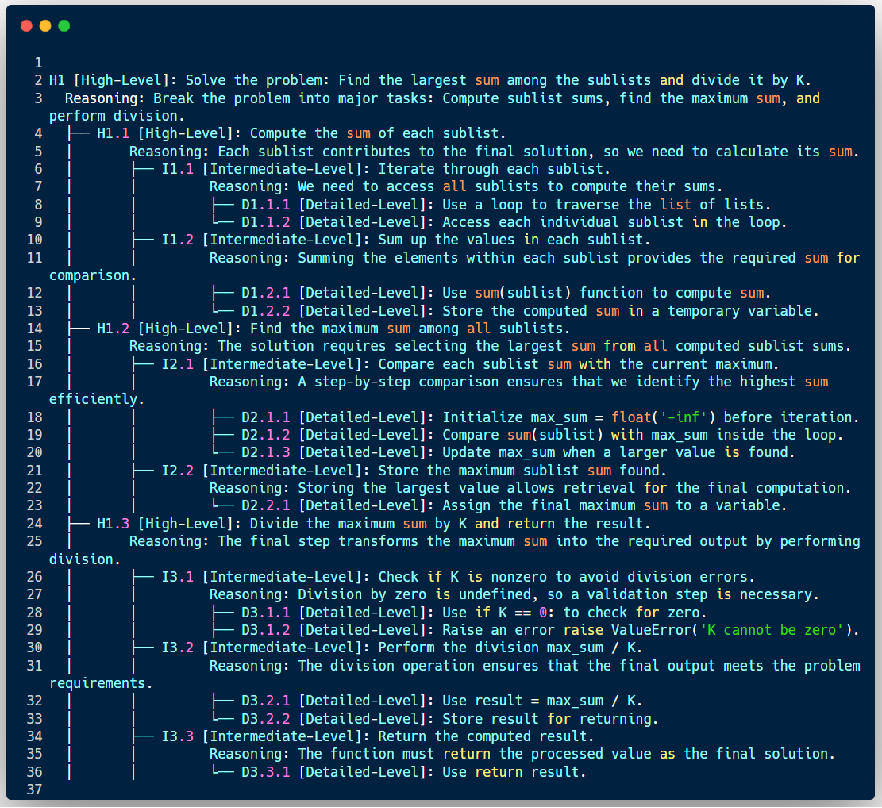}
\caption{Generated MLR Graph illustrating modular reasoning hierarchy.}
\end{figure}

\begin{figure}[t]
\centering
\begin{minipage}{0.5\textwidth}
\begin{lstlisting}[style=promptstyle]
def validate_input(K):  
    """D3.1.1: Check if K is zero"""
    if K == 0:
        raise ValueError("K cannot be zero")  # D3.1.2: Raise error for zero division

def compute_sublist_sum(sublist):
    """D1.2.1: Compute sum using sum() function"""
    return sum(sublist)

def find_max_sublist_sum(array):
    """D2.1.1: Initialize max sum and find the maximum sublist sum"""
    max_sum = float('-inf')
    for sublist in array:
        sublist_sum = compute_sublist_sum(sublist)
        if sublist_sum > max_sum:
            max_sum = sublist_sum
    return max_sum

def divide_max_sum(max_sum, K):
    """D3.2.1: Perform division"""
    return max_sum / K

def largest_sum(array, K):
    """Main function that orchestrates modularized components"""
    validate_input(K)
    max_sum = find_max_sublist_sum(array)
    return divide_max_sum(max_sum, K)
\end{lstlisting}
\captionsetup{type=figure}
\caption{Generated Modular Code based on the MLR Graph}
\end{minipage}
\end{figure}